\documentclass[aps,prl,twocolumn,superscriptaddress,letterpaper,floatfix]{revtex4-1}
\usepackage{graphicx,color}
\usepackage{verbatim}
\usepackage{amssymb}   
\usepackage{amsmath}
\usepackage{amsfonts}
\usepackage{mathdots}
\usepackage{hyperref}
\usepackage{epsfig}

\usepackage{braket}
\usepackage{bm}
\usepackage{physics}

\begin{document}

\title{Flat-band Fulde-Ferrell-Larkin-Ovchinnikov State from Quantum Geometric Discrepancy}

\author{Zi-Ting Sun}\thanks{These authors contributed equally to this work.}
\author{Ruo-Peng Yu}\thanks{These authors contributed equally to this work.}

\affiliation{Department of Physics, Hong Kong University of Science and Technology, Clear Water Bay, Hong Kong, China}
\author{Shuai A. Chen}
\affiliation{Max Planck Institute for the Physics of Complex Systems, N\"{o}thnitzer Stra{\ss}e 38, Dresden 01187, Germany}

\author{Jin-Xin Hu}\thanks{Contact author: jhuphy@ust.hk}
\affiliation{Division of Physics and Applied Physics, School of Physical and Mathematical Sciences, Nanyang Technological University, Singapore 637371} 

\author{K. T. Law}\thanks{Contact author: phlaw@ust.hk}
\affiliation{Department of Physics, Hong Kong University of Science and Technology, Clear Water Bay, Hong Kong, China}

	\date{\today}
	\begin{abstract}
We propose a new scheme for realizing Fulde-Ferrell-Larkin-Ovchinnikov (FFLO) Cooper pairing states within flat bands, in contrast to the conventional paradigm such as the Zeeman effect. Central to our scheme is the concept of ``quantum geometric discrepancy'' (QGD) that measures differences in the quantum geometry of paired electrons and drives the flat-band FFLO instability. Remarkably, we find that this instability is directly related to a quantum geometric quantity known as ``anomalous quantum distance'', which formally captures QGD. To model both QGD and the anomalous quantum distance, we examine a flat-band electronic Hamiltonian with tunable spin-dependent quantum metrics. Utilizing the band-projection method, we analyze the QGD-induced FFLO instability from pairing susceptibility. Furthermore, we perform mean-field numerical simulations to obtain the phase diagram of the BCS-FFLO transition, which aligns well with our analytical results. Our work demonstrates that QGD offers a general and distinctive mechanism for stabilizing the flat-band FFLO phase.

	\end{abstract}
	\maketitle

\section{Introduction}
Understanding unconventional superconductivity is a central theme in modern condensed matter physics, with important implications for both theoretical and experimental aspects~\cite{sigrist1991phenomenological,mineev1999introduction,bauer2012non,scalapino2012common}. Among the numerous mechanisms extending beyond the conventional Bardeen-Cooper-Schrieffer (BCS) theory of pairing, the Fulde-Ferrell-Larkin-Ovchinnikov (FFLO) state stands out due to its characteristic finite center-of-mass momentum Cooper pairing~\cite{fulde1964superconductivity,larkin1965nonuniform,kinnunen2018fulde,agterberg2020physics}. Various mechanisms have been explored to realize exotic FFLO states in systems characterized by dispersive bands~\cite{buzdin1997generalized,liu2017unconventional,song2019quantum,xie2023orbital,yuan2023orbital,nakamura2024orbital,wan2023orbital}. 

\begin{figure}
		\centering
		\includegraphics[width=1.0\linewidth]{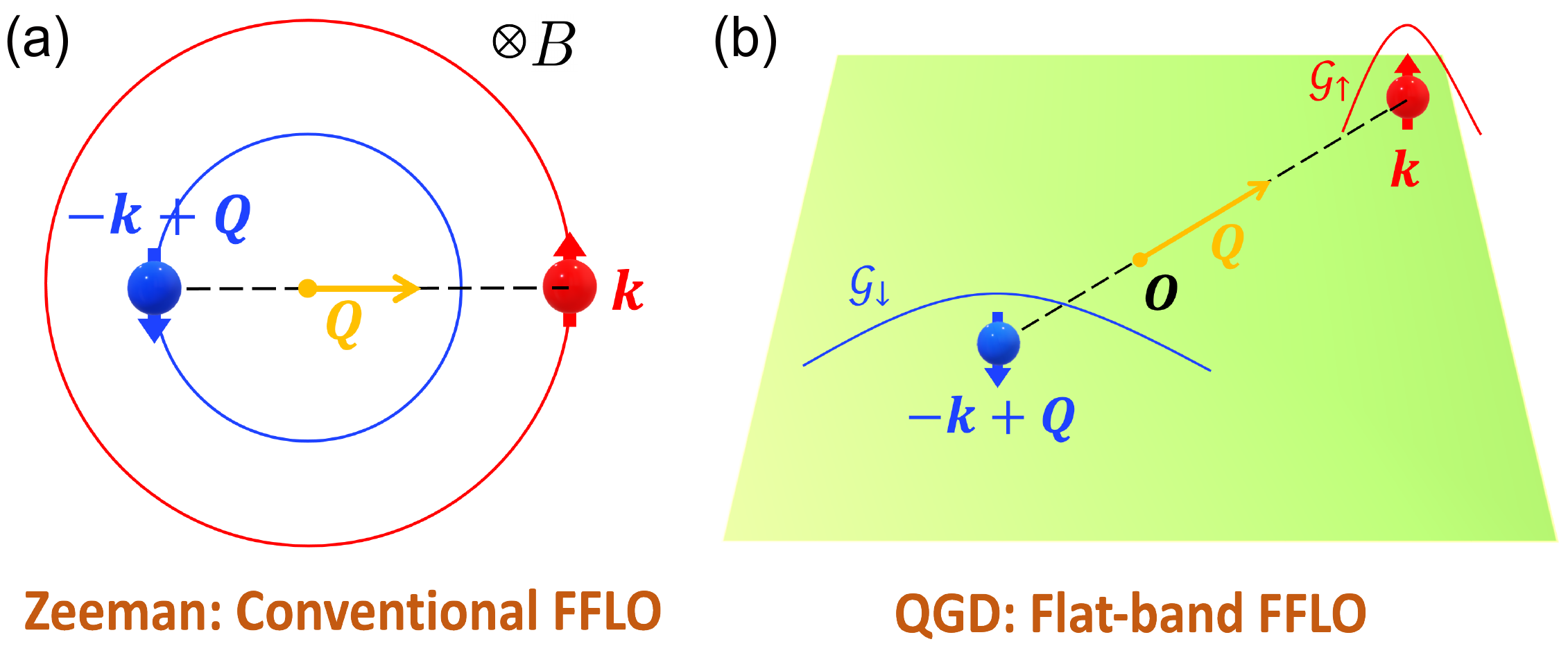}
		\caption{Schematic illustrations of (a) conventional FFLO from the Zeeman effect, and (b) flat-band FFLO from QGD. In (a), an external $B$ field induces spin-split Fermi surfaces,  leading to the formation of Cooper pairs with finite momentum $\bm{Q}$, where $|\bm{Q}| \propto B/v_F$. In (b), the finite momentum $\bm{Q}$ is stabilized when the quantum metrics of the paired electrons are different ($\mathcal{G}_\uparrow \neq \mathcal{G}_\downarrow$) within flat bands.} 
		\label{fig:fig1}
\end{figure}

In the original FFLO formulation, a spin-singlet pairing state develops finite center-of-mass momentum $\bm{Q}$ ($|\bm{Q}|\propto B/v_F$) under a strong Zeeman field $B$ [see Fig.~\ref{fig:fig1}(a)]. However, this microscopic mechanism becomes problematic in flat-band systems due to the vanishing Fermi velocity ($v_F\rightarrow 0$), which would lead to an unphysical divergence of the pairing momentum $\bm{Q}$. Consequently, two fundamental questions arise: First, what mechanism can stabilize the FFLO state in the flat-band limit ($v_F\rightarrow 0$)? Second, how is the center-of-mass momentum $\bm{Q}$ of Cooper pairs determined in flat bands? While Fermi surface nesting typically governs $\bm{Q}$ selection in dispersive band systems, this relationship becomes ambiguous in flat bands due to the ill-defined Fermi surface.

Building upon recent studies of the quantum geometric effects in flat-band systems~\cite{peotta2015superfluidity,julku2016geometric,liang2017band,torma2018quantum,hu2019geometric,julku2020superfluid,xie2020topology,peri2021fragile,he2021geometry,verma2021optical,herzog2022superfluid,torma2022superconductivity,huhtinen2022revisiting,kitamura2022quantum,hofmann2023superconductivity,jiang2023pair,chen2023pair,torma2023essay,tian2023evidence,han2024quantum,ticea2024pair,wang2024quantum,chen2024ginzburg,yu2024non,yu2024quantum,hu2025anomalous},  we address the aforementioned questions by linking the FFLO instability to the quantum geometry of Bloch wave functions. For concreteness, we propose a new scheme for FFLO pairing states within flat bands, driven by ``quantum geometric discrepancy" (QGD). This concept refers to the differences in the quantum geometry of the wave functions of the two electrons forming a Cooper pair. Based on the band-projection formalism, we analyze the static pairing susceptibility and construct that the critical temperature for finite-momentum pairing, which leads to FFLO instability, is directly associated with the band average of a quantum geometric quantity [see Eq.~\eqref{critem}]—“anomalous quantum distance" (AQD) $\bar{d}_{\bm{k}',\bm{k}}$, formally defined as
\begin{equation} \label{eq:aqd} 
\bar{d}^2_{\bm{k}',\bm{k}}=1-\left| \langle u_{\bm{k}'\uparrow}|u^*_{-\bm{k}\downarrow}\rangle\right|^2. \end{equation}  
Here, $u_{\bm{k}\sigma}$ represents the lattice-periodic part of the Bloch wave function at momentum $\bm{k}$ and spin $\sigma$. AQD quantifies the overlap between the wave functions of paired electrons. In the presence of time-reversal ($\mathcal{T}$) symmetry, where $u_{\bm{k}\uparrow}=u^*_{-\bm{k}\downarrow}$, the quantity $\bar{d}_{\bm{k},\bm{k}'}$ becomes the well-known Hilbert–Schmidt quantum distance \cite{shapere1989geometric,buvzek1996quantum,dodonov2000hilbert,rhim2020quantum}, measuring the quantum mechanical distance between two Bloch states. In this scenario, the BCS (zero-momentum) pairing state is favored. In the absence of $\mathcal{T}$ symmetry, $u_{\bm{k}\uparrow}\neq u^*_{-\bm{k}\downarrow}$ generally makes this conclusion invalid, and Eq.~\eqref{eq:aqd} will be frustrated due to the QGD, enabling a finite-momentum pairing state. 

As we will substantiate, even in the flat-band limit, a finite momentum-$\bm{Q}$ pairing state can be stabilized when the pairing is frustrated by QGD, specifically as $\mathcal{G}_{\uparrow}\neq \mathcal{G}_{\downarrow}$ [see Fig.~\ref{fig:fig1}(b)]. Here, $\mathcal{G}_{\uparrow}$ and $\mathcal{G}_{\downarrow}$ represent the quantum metrics of the Bloch states for the paired electrons. Our work establishes a comprehensive theoretical framework for understanding FFLO states in flat-band systems.

\section{Effective action from band projection}
To investigate the FFLO instability on the flat band, we first derive the superconducting Gaussian fluctuations using the band-projection formalism~\cite{iskin2023extracting,chen2024ginzburg}, and this analysis is rather general when considering uniform pairing. Without loss of generality, we consider the local attractive Hubbard interaction $H_{\mathrm{int}}=-U\sum_{i\alpha}\hat{c}^{\dagger}_{i\alpha \uparrow}\hat{c}^{\dagger}_{i\alpha \downarrow}\hat{c}_{i\alpha \downarrow}\hat{c}_{i\alpha \uparrow}$, where $U>0$, $\alpha$ and $i$ denote the orbital and site index. 

In order to analyze the low-energy behavior of superconductivity on the relevant isolated band using field theory, we first implement the Fourier transformation ($\hat{c}_{i\alpha \sigma}=1/\sqrt{N_c}\sum_{\bm{k}} e^{i \bm{k} \cdot \bm{r}_{i}} \hat{c}_{\alpha, \bm{k} \sigma}$, where $N_c$ is the number of unit cells) and the uniform pairing approximation~\cite{torma2018quantum,huhtinen2022revisiting}, then we project the interaction onto the target band by $\hat{c}_{\alpha, \boldsymbol{k} \sigma} \rightarrow u_{\boldsymbol{k}\sigma}^*(\alpha) \hat{a}_{\boldsymbol{k}\sigma}$, where $\hat{c}_{\alpha, \boldsymbol{k} \sigma}$ and $\hat{a}_{\boldsymbol{k}\sigma}$ denotes the orbital and band-basis electron annihilation operators. We can obtain the effective attractive interaction
\begin{equation}\label{interact}
\begin{aligned}
&H_{\mathrm{int}}=-g\sum_{\boldsymbol{q}}\hat{\theta}^{\dagger}_{\bm{q}}\hat{\theta}_{\bm{q}},\\
\hat{\theta}_{\bm{q}}=&\sum_{\boldsymbol{k}}\Lambda^*(\boldsymbol{k},\boldsymbol{q})\hat{a}_{-\boldsymbol{k}\downarrow}\hat{a}_{\boldsymbol{k}+\boldsymbol{q}\uparrow},
\end{aligned}
\end{equation}
in which $\Lambda(\bm{k},\bm{q})=\sum_{\alpha}u_{\boldsymbol{k}+\boldsymbol{q}\uparrow}(\alpha)u_{-\boldsymbol{k}\downarrow}(\alpha)$ is the form factor, which encodes the information of quantum geometry. Here the effective coupling constant $g\approx U/N_{\mathrm{orb}}N_c$, where $N_{\mathrm{orb}}$ is the number of orbitals in each unit cell. For convenience, we naturally introduce the narrow band condition such that the attractive interaction strength $g$ is large compared to the bandwidth $W$ of the isolated flat band but small compared to the band gap $\Delta_{\mathrm{gap}}$ ($W\ll g\ll \Delta_{\mathrm{gap}}$). From Eq.~\eqref{interact}, we can arrive at an effective one-band description in the functional integral formalism~\cite{setty2023mechanism,Altland_Simons_2023}, and the effective action can be expressed as
\begin{equation}\label{action}
  S=\sum_{k, \sigma} \bar{a}_{k \sigma}\left(-i \omega_n+\xi_{\bm{k}\sigma}\right) a_{k \sigma}-g\sum_{q}\bar{\theta}_{q}\theta_{q}.
\end{equation}
 Here we have used the notations $k=\left(\boldsymbol{k}, \omega_n\right)$ and $q=\left(\boldsymbol{q}, \Omega_m\right)$ with $\omega_n$ $(\Omega_m)$ being the fermionic (bosonic) Matsubara frequencies, and $\xi_{\bm{k}\sigma}$ is the band dispersion for spin-$\sigma$ electrons. Then we adopt the standard Hubbard-Stratonovich decoupling of Eq.~\eqref{action} to derive the effective action $S[\Delta_q]$ (for details see Supplementary Information (SI) I.A~\cite{supp}) of the superconducting order parameter field $\Delta_{q}$. By retaining the second order terms of $\Delta_{q}$ in $S[\Delta_q]$, the Gaussian action around the normal state saddle point $\Delta_q=0$ reads
\begin{equation}
S_G\left[\Delta_{q}\right]=S[\Delta_q=0]+\sum_{q} \Gamma_{q}^{-1}|\Delta_{q}|^2,
\end{equation}
where the coefficient $\Gamma_{q}^{-1}=g^{-1}-\chi^c_{q}$ is the Cooper-pair propagator, and the static pairing susceptibility $\chi^c_{\boldsymbol{q}}\equiv \chi^c_{q=(\boldsymbol{q},0)}$ reads
\begin{equation}\label{sus}
\chi^c_{\boldsymbol{q}}=\frac{1}{N_c} \sum_{\boldsymbol{k}} \frac{1-n_{\mathrm{F}}\left(\xi_{\boldsymbol{k+q}}\right)-n_{\mathrm{F}}\left(\xi_{\boldsymbol{k}}\right)}{\xi_{\boldsymbol{k+q}}+\xi_{\boldsymbol{k}}}|\Lambda(\boldsymbol{k}, \boldsymbol{q})|^2.
\end{equation}
Here we use the notation as $\xi_{\bm{k}}=\xi_{\bm{k} \uparrow}=\xi_{-\boldsymbol{k} \downarrow}$ and $n_{\mathrm{F}}$ is the Fermi-Dirac distribution. The number equation near the critical temperature is $N=-\beta^{-1}\partial S[\Delta_q=0]/\partial \mu$, where $N$ is the total particle number. This yields $2\mathcal{A}\nu=\sum_{\boldsymbol{k}}\left[1-\tanh \left(\beta \xi_{\boldsymbol{k}}/2\right)\right]$, where $\mathcal{A}$ is the area of the first Brillouin zone, $\beta^{-1}=k_B T$, and $\nu$ is the filling factor of the target band ($0\leq\nu\leq1$).

\section{Critical temperature of FFLO instability}
By solving the linearized gap equation $g\chi^c_{\boldsymbol{q}}=1$ together with the number equation, we can derive the critical temperature $T_{c\boldsymbol{q}}$ for the instability of a static, spatially non-uniform pairing state characterized by a finite $\boldsymbol{q}$ index. Restricting the analysis to single-$\boldsymbol{q}$ states, $\Delta_{\boldsymbol{q}} \propto \delta_{\boldsymbol{q}, \boldsymbol{Q}}$, the value of $\boldsymbol{Q}$ is found at the highest $T_{c\boldsymbol{q}}$ for different $\boldsymbol{q}$. If $\bm{Q}\neq 0$ ($\bm{Q}= 0$), the pairing instability is associated with the FFLO (BCS) phase.

Importantly, in the exact flat-band limit, we find $\chi^c_{\boldsymbol{q}}=  \beta\sum_{\boldsymbol{k}}|\Lambda(\boldsymbol{k},\boldsymbol{q})|^2 (1/2-\nu)/[N_c \ln(\nu^{-1}-1)]$. Rewriting Eq.~\eqref{eq:aqd} as $\bar{d}^2_{\boldsymbol{k}+\boldsymbol{q},\boldsymbol{k}}=1-|\Lambda(\boldsymbol{k},\boldsymbol{q})|^2$, $T_{c\boldsymbol{q}}$ with fixed band filling $\nu$ and coupling constant $g$ can be derived as
\begin{equation}\label{critem}
   T_{c\boldsymbol{q}}/T_0=1-\bar{d}^2_{\mathcal{A}}(\boldsymbol{q}),
\end{equation}
in which the average of the AQD over the Brillouin zone within the band is defined as $\bar{d}^2_{\mathcal{A}}(\boldsymbol{q})=\sum_{\boldsymbol{k}} \bar{d}^2_{\boldsymbol{k}+\boldsymbol{q},\boldsymbol{k}}/N_c$, and $T_0=g(1/2-\nu)/ [\ln(\nu^{-1}-1)]$ is a constant which is independent of $\boldsymbol{q}$. Notice that $T_0\propto g$, which gets maximized at $\nu\rightarrow 1/2$ and vanishes at $\nu\rightarrow 0$ and $\nu\rightarrow 1$, so does $T_{c\boldsymbol{q}}$. When we fix $T_0$, the highest critical temperature should be determined by $T_{c\boldsymbol{Q}}=T_0\{1-\mathrm{min}[\bar{d}^2_{\mathcal{A}}(\bm{q})]\}$ over different $\boldsymbol{q}$ in the first Brillouin zone, and the corresponding optimal momentum is $\boldsymbol{q}=\boldsymbol{Q}$, which indicates the flat-band FFLO instability is solely determined by the quantum geometric quantity $\bar{d}^2_{\mathcal{A}}(\bm{q})$. This is the main result of our work. Although Eq.~\eqref{critem} is derived neglecting fluctuations, we argue that after considering the corrections of number equation in the presence of Gaussian fluctuations, the optimal momentum with highest critical temperature can still be determined solely by AQD, and the emergence of FFLO instability on the whole aligns with the predictions from Eq.~\eqref{critem}, see SI I.B~\cite{supp} for more details.

For a $\mathcal{T}$-symmetric system, $u_{\bm{k}\uparrow} = u^*_{-\bm{k}\downarrow}$ renders $\bar{d}_{\bm{k}+\boldsymbol{q},\bm{k}}$ being the ordinary quantum distance. Since $\bar{d}^2_{\bm{k},\bm{k}} = 0$, the quantity $\bar{d}^2_{\mathcal{A}}(\bm{q})$ possesses a global minimum at $\bm{q} = 0$. As we will see in the following sections, the introduction of QGD explicitly breaks the $\mathcal{T}$ symmetry and potentially shifts the global minimum of $\bar{d}^2_{\mathcal{A}}(\bm{q})$ away from $\bm{q} = 0$, implying the BCS to FFLO transition. Thus, with Eq.~\eqref{critem} in hand, the concept of QGD is natural in stabilizing the flat-band FFLO state.

\begin{figure}
		\centering
		\includegraphics[width=1\linewidth]{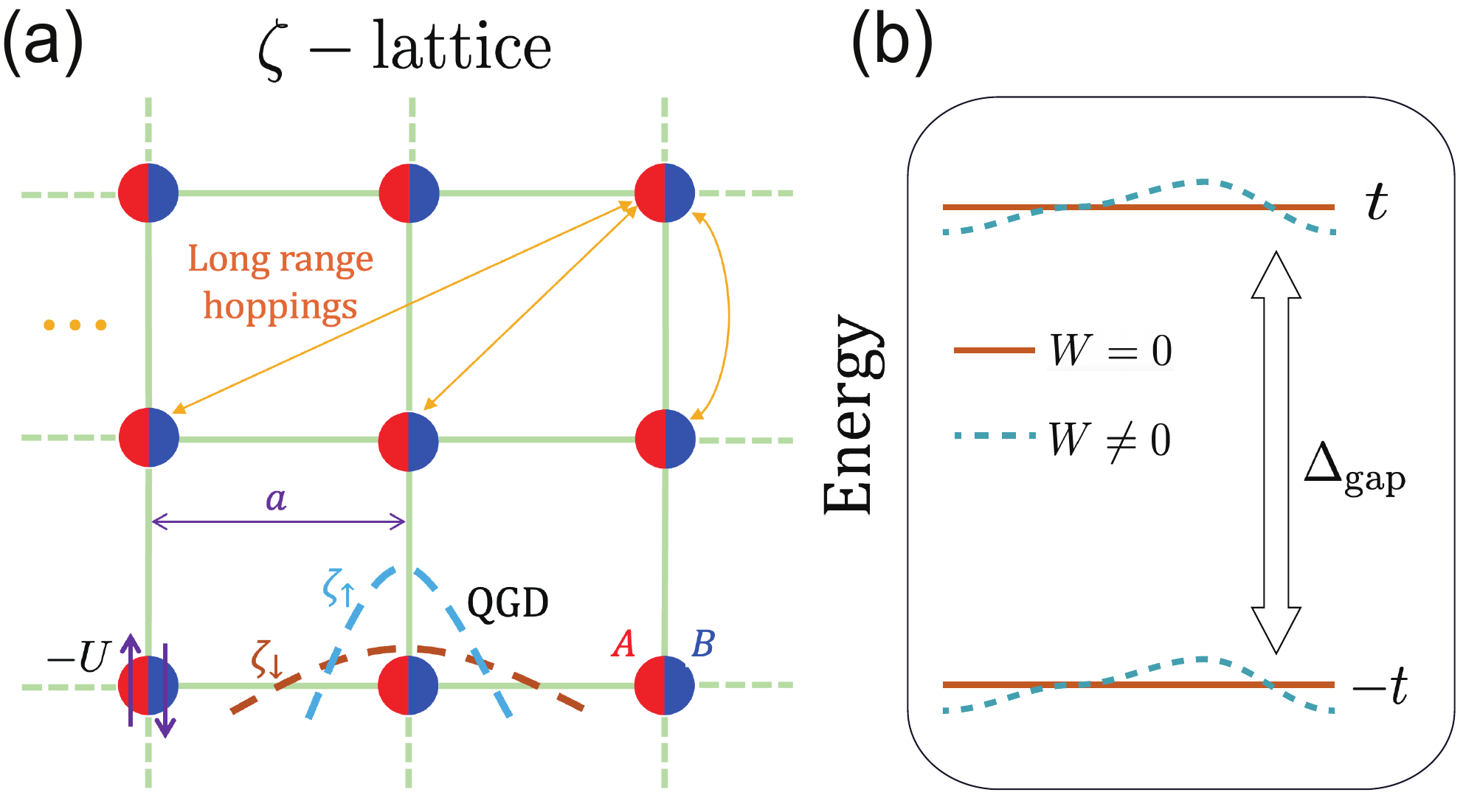}
		\caption{(a) The $\zeta$-lattice with spin-dependent long-range hoppings. $\zeta_{\sigma}$ denotes the quantum metric of spin-$\sigma$ sector. (b) Energy dispersions of the $\zeta$-lattice model in flat-band limit ($W=0$) and with finite bandwidth ($W\neq 0$). Band gap $\Delta_{\mathrm{gap}}$ is much lager than bandwidth $W$.}
		\label{fig:fig2}
\end{figure}

\section{Model Hamiltonian}
To exemplify the Cooper pairings with QGD, we analyze a concrete model~\cite{hofmann2022heuristic, hofmann2023superconductivity, mao2023diamagnetic} that allows for continuous tuning of the quantum geometry. As shown in Fig.~\ref{fig:fig2}(a), the $\zeta$-lattice model features two equivalent orbitals (A and B) per site on a square lattice, described by the Hamiltonian $H=\sum_{\boldsymbol{k} \sigma} \hat{c}_{\boldsymbol{k} \sigma}^{\dagger} (\mathcal{H}_{\boldsymbol{k} \sigma} -\mu_{\sigma})\hat{c}_{\boldsymbol{k} \sigma}$, where
\begin{equation}
\label{eq:hamiltonian}
\mathcal{H}_{\boldsymbol{k} \sigma}(\zeta_\sigma)=-t \left[\lambda_x \sin (\zeta_\sigma p_{\bm{k}})  +m_\sigma \lambda_y \cos(\zeta_\sigma p_{\bm{k}})\right],
\end{equation}
with $p_{\bm{k}}=\cos k_x +\cos k_y$. For brevity we set the lattice constant $a=1$ and $\mu_\sigma=\mu$. With the spin index $\sigma=\uparrow, \downarrow$, the fermion operators are defined as $\hat{c}_{\boldsymbol{k} \sigma}=(\hat{c}_{A,\boldsymbol{k} \sigma},\hat{c}_{B ,\boldsymbol{k} \sigma})^T$. The Pauli matrices $\lambda_i$ act on orbital space, and time-reversal flavors are related by $m_{\sigma=\uparrow/\downarrow}= \pm 1$. The periodic function $\zeta_\sigma p_{\bm{k}}$ contains spin-dependent long range hoppings [see Fig.~\ref{fig:fig2}(a)]. As plotted in Fig.~\ref{fig:fig2}(b), this model Hamiltonian possesses a pair of ideal isolated flat bands with band gap $\Delta_{\mathrm{gap}}=2t$. For each spin flavor, the band dispersion $\xi_{\boldsymbol{k}\sigma}$ as well as Bloch function $u_{\bm{k}\sigma}$ can be solved as $\xi_{\boldsymbol{k}\sigma}= \pm t-\mu$ and 
\begin{equation}
u_{\bm{k}\sigma}=\frac{1}{\sqrt{2}}\begin{pmatrix}
          \pm 1 \\
          e^{i m_\sigma \zeta_{\sigma} p_{\bm{k}} }     
         \end{pmatrix}.
\end{equation}
Given the Bloch function $u_{\bm{k}\sigma}$, the spin-dependent quantum metrics can be obtained as $\mathcal{G}_{ab}(\bm{k},\sigma) = \zeta_{\sigma}^2\sin(k_a) \sin(k_b)/4$. 

When considering electrons with opposite momentum $\bm{k}$ that yield spin-singlet pairing, the time-reversal partners $(\bm{k},\uparrow)$ and $(-\bm{k},\downarrow)$ form a Cooper pair. Crucially, $(\bm{k},\uparrow)$ and $(- \bm{k},\downarrow)$ are degenerate in energy as $\xi_{\bm{k}\uparrow} = \xi_{-\bm{k}\downarrow}$, but are polarized in quantum metrics when $\zeta_{\uparrow} \neq \zeta_{\downarrow}$, meaning $\mathcal{G}(\bm{k},\uparrow) \neq \mathcal{G}(-\bm{k},\downarrow)$. In this case, $\mathcal{T}$ symmetry of the system is broken, which is evidenced by $u_{\bm{k}\uparrow} \neq u_{-\bm{k}\downarrow}^*$, rather than by energy difference. For the $\zeta$-lattice model, we define the dimensionless parameter $\eta = |\zeta_{\uparrow} - \zeta_{\downarrow}|/(\zeta_{\uparrow} + \zeta_{\downarrow})$ as a global measure of QGD. In the following, we will illustrate the role of QGD in the formation of flat-band FFLO states.

\section{QGD-stabilized flat-band FFLO state}
\begin{figure}[t]
		\centering
		\includegraphics[width=1\linewidth]{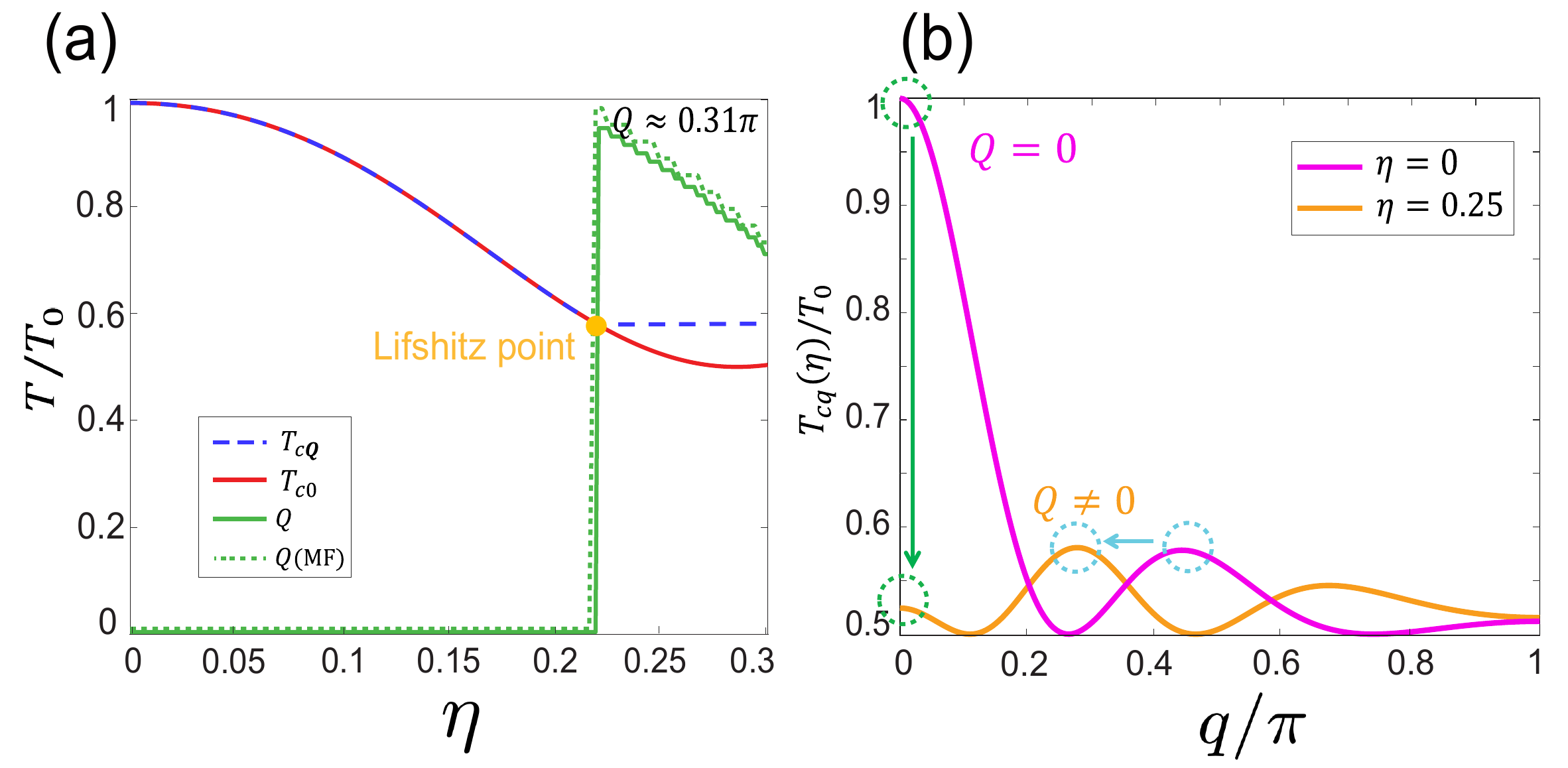}
		\caption{(a) $T_{c\bm{Q}}(\eta)$ (dashed blue), $T_{c0}(\eta )$ (solid red), and the corresponding $Q(\eta )$ (solid green) calculated from Eq.~\eqref{critem} as a funtion of $\eta$. The orange dot labels the Lifshitz point. For comparison, the dashed green line shows $Q(\eta)$ calculated from the self-consistent mean-field (MF) theory. (b) $T_{c\bm{q}}(\eta )/T_0$ from Eq.~\eqref{critem} for $\boldsymbol{q} = q(\hat{x}+\hat{y}),\, q\in \left [ 0,\pi  \right ] $, at $\eta=0$ and $\eta=0.25$, respectively. In our calculations, we set $\zeta_{\downarrow}=3$ and $\zeta_{\uparrow}=\zeta_{\downarrow}(1+\eta)/(1-\eta)$.}
		\label{fig:fig3}
\end{figure}

In terms of Eq.~\eqref{critem}, in Fig.~\ref{fig:fig3}(a) we plot $T_{c\boldsymbol{Q}}(\eta)$ (dashed blue curve) as well as the corresponding $\boldsymbol{Q}(\eta )$ (solid green curve). As a comparison, $T_{c0}(\eta )$ (solid red curve) is also depicted. In the BCS phase where $\eta<\eta_c=0.22$, we observe $T_{c\boldsymbol{Q}}(\eta)=T_{c0}(\eta )$ with $\boldsymbol{Q}(\eta )=0$. For $\eta>\eta_c$, however, $T_{c\boldsymbol{Q}}(\eta)> T_{c0}(\eta)$, and $\boldsymbol{Q}(\eta )$ undergoes a discontinuous jump from $Q(\eta_c-)=0$ to $Q(\eta_c+)\approx0.31\pi$, signaling a first-order BCS to FFLO phase transition. The Lifshitz point (orange dot) at $(\eta_c=0.22, T=0.57T_0)$ marks the tri-critical point where the BCS, FFLO, and normal phases coexist. We also find that $\bm{Q}$ is four-fold degenerate along $\pm Q(\hat{x}+\hat{y})$ and $\pm Q(-\hat{x}+\hat{y})$ (see SI III.A~\cite{supp}); throughout this work we take $\bm{Q}=Q(\hat{x}+\hat{y})$ for simplicity.

To further understand the QGD-driven FFLO transition and the role of AQD, we plot the $T_{c\boldsymbol{q}}(\eta)$ evaluated from Eq.~\eqref{critem} in Fig.~\ref{fig:fig3}(b), for $\eta=0$ (pink curve) and $\eta=0.25$ (orange curve), respectively. The highest critical temperature is determined by the minimum of $\bar{d}^2_{\mathcal{A}}(\bm{q})$, which corresponds to the maximum of $T_{c\boldsymbol{q}}$. As we mentioned before, $\mathcal{T}$ symmetry ($\eta=0$) ensures the favorable BCS pairing state ($Q=0$) because the global maximum of $T_{c\boldsymbol{q}}$ always occurs at $q=0$ (dashed green circle on the pink curve). Note that a local maximum of $T_{c\boldsymbol{q}}(0)$ can be seen at nonzero $q$ in Fig.~\ref{fig:fig3}(b). As QGD ($\eta \ne0$) decreases $T_{c0}(\eta)$ significantly (the green dashed circle on the orange curve), it is possible that a nonzero $Q$ can be stabilized. Upon increasing $\eta$, the local maximum finally transitions to global maxima (the cyan dashed circle on the orange curve) at $Q(\eta )\ne0$, where a FFLO phase can naturally manifest. Here our calculations are based on the square lattice model, and in SI III.A~\cite{supp} we also examine a triangular lattice model, where our theory still applies. Our analysis based on pairing susceptibility is supported by a recent follow-up work \cite{zhang2025identifying} through numerical determinantal quantum Monte Carlo calculations.

\section{Self-consistent mean field calculations}
We have seen that the pairing susceptibility analysis, including Eq.~\eqref{sus} and \eqref{critem}, have a profound impact on identifying FFLO instabilities within the single-band description. To systematically verify these findings and get a full phase diagram, we perform numerical simulations using self-consistent mean-field theory~\cite{kitamura2022quantum}. These calculations clearly demonstrate a first-order BCS to FFLO phase transition, directly driven by QGD in our model Hamiltonian [Eq.~\eqref{eq:hamiltonian}]. For clarity, the mean-field decoupling of the attractive Hubbard interaction yields orbital-dependent order parameters as $\Delta_{i\alpha }=-U\langle  \hat{c}_{i \alpha \downarrow}\hat{c}_{i \alpha \uparrow} \rangle=\Delta_{\alpha  }(\bm{q})e^{i \bm{q}\cdot  \bm{r}_i}$ with $\bm{q}$ index. In the canonical ensemble, the Helmholtz free energy density $F$ can be written as
\begin{equation}\label{free}
\begin{aligned}
    F(\bm{q}) =& - \frac{1}{\beta N_c}\sum_{\bm{k}}\operatorname{Tr}\left[\operatorname{ln}\left(1 + e^{-\beta H_{\mathrm{BdG}}(\bm{k},\bm{q})}\right)\right] \\
    &+\frac{1}{U}\sum_{\alpha}|\Delta_{\alpha}(\bm{q})|^2,
\end{aligned}
\end{equation}
where the Bogoliubov–de Gennes (BdG) Hamiltonian $H_{\mathrm{BdG}}(\bm{k},\bm{q})$ reads
\begin{equation}
H_{\mathrm{BdG}}(\bm{k},\bm{q}) = \left(\begin{array}{cc}
         \mathcal{H}_{\bm{k}+\bm{q}\uparrow} - \mu_{\bm{q}} & \hat{\Delta}_{\bm{q}} \\
        \hat{\Delta}^{\dagger}_{\bm{q}} & 
         -\mathcal{H}_{-\bm{k}\downarrow}^{*} + \mu_{\bm{q}}
    \end{array}
    \right),
\end{equation}
with the order parameters $\hat{\Delta}_{\bm{q}} = \mathrm{diag}[\Delta_{A}(\bm{q}),\Delta_{B}(\bm{q})]$. $\hat{\Delta}_{\bm{q}}$ and $\mu_{\bm{q}}$ can be determined self-consistently by solving the gap and number equations, respectively. Substituting $\hat{\Delta}_{\bm{q}}$ and $\mu_{\bm{q}}$ into Eq.~\eqref{free}, we can obtain $F(\boldsymbol{q})$ for each $\bm{q}$ (see more details in SI II~\cite{supp}).

\begin{figure}[b]
        \centering
		\includegraphics[width=1.0\linewidth]{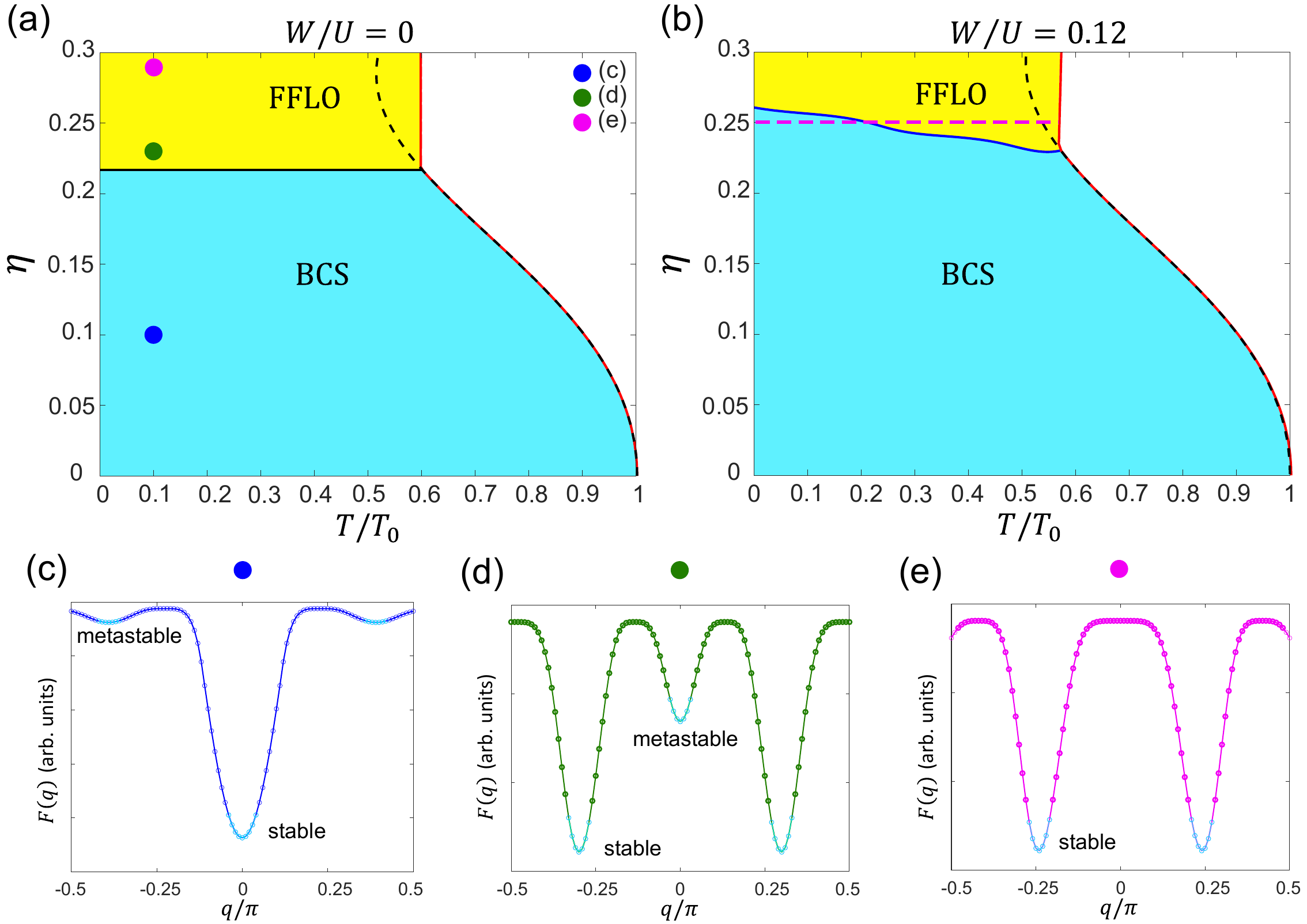}
		\caption{The BCS-FFLO phase diagram, without (a) and with (b) band dispersion. The dashed black line labels $T_{c0}(\eta)$, while the solid red line labels $T_{c\boldsymbol{Q}}(\eta)$. The BCS and FFLO phases are colored in light blue and yellow, respectively. Parameters used here: $U = 0.4t$, $\nu = 0.5$, $\zeta_{\downarrow} = 3$, $W/U=0$ in (a) and $W/U=0.12$ in (b). (c) (d) (e): The free energy density landscapes of three highlighted dots in (a) at $T = 0.1T_0$ and different $\eta$ ($\eta = 0.10$, $0.23$, and $0.29$, respectively).}
		\label{fig:fig4}
\end{figure}

The numerical results are summarized in Fig.~\ref{fig:fig4}. As shown in Fig.~\ref{fig:fig4}(a), the BCS and FFLO phases are colored light blue and yellow, respectively. The critical temperatures $T_{c0}(\eta)$ as well as $T_{c\boldsymbol{Q}}(\eta)$ are plotted in the dashed black and the solid red curves. We find that the optimal pairing momentum $\boldsymbol{Q}(\eta)$ is independent of the temperature, resulting in a horizontal boundary line between the BCS and FFLO phases. In this case, the thermal fluctuations play a negligible role and the phase transition is solely driven by QGD. The calculated $T_{c0}(\eta)$, $T_{c\boldsymbol{Q}}(\eta)$ and $\bm{Q}(\eta)$ exhibit quantitative agreement with the corresponding analytical results in Fig.~\ref{fig:fig3}(a).  

To elucidate the nature of the phase transition, in Fig.~\ref{fig:fig4}(c) to (e), we show the landscapes of free energy $F(\boldsymbol{q})$ at the three points (blue, green, and pink dots) in Fig.~\ref{fig:fig4}(a). Initially, in the BCS region ($\eta<\eta_c$) [Fig.~\ref{fig:fig4}(c)], the global minimum of the free energy $F(\bm{q})$ is at $\bm{q}=0$, which is stable under weak perturbations. Some metastable states exist at $\bm{q}\neq0$ as local minima in $F(\bm{q})$. When $\eta >\eta_c$ [Fig.~\ref{fig:fig4}(d)], $\bm{q}=0$ turns into a local minimum but remains a metastable state, while some $\bm{q}=\bm{Q}$ points become global minima (the model respects $C_4$ rotational symmetry, so the minima have four-fold degeneracy in the whole first Brillouin zone, see SI III.A~\cite{supp}), the FFLO phase can manifest. This indicates that the QGD energetically penalizes BCS pairing more than FFLO pairing. A first-order phase transition from BCS to FFLO occurs near $\eta_c$, which is consistent with the results obtained from the pairing susceptibility. As $\eta$ continues to increase [Fig.~\ref{fig:fig4}(e)], the local minimum at $\bm{q} = 0$ eventually nearly disappears.

\section{Robustness with finite bandwidth}
The QGD-driven BCS-FFLO phase transition has been examined in the perfect flat-band limit, both analytically and numerically. To show the robustness of QGD-induced BCS-FFLO transition with finite band dispersion, we modify the Bloch Hamiltonian in Eq.~\eqref{eq:hamiltonian} as $\mathcal{H}'_{\bm{k} \sigma}=\mathcal{H}_{\bm{k} \sigma}-W\left(\cos k_x+\cos k_y\right)\lambda_0/4$, where an additional nearest hopping term brings a finite bandwidth $W$ as shown in Fig.~\ref{fig:fig2}(b). In this case, the Bloch wave functions remain unchanged. We keep $W\ll U$ to ensure a narrow bandwidth compared to the interaction. 

In Fig.~\ref{fig:fig4}(b), the mean-field phase diagram at $W/U=0.12$ ($U=0.4t$) is shown. Self-evidently, compared with Fig.~\ref{fig:fig4}(a) ($W/U=0$), the transition boundary line is temperature-dependent and becomes curved due to the nonzero $W$, requiring a larger QGD to reach the FFLO region, especially at lower temperatures. This fact indicates that increasing the temperature may also drive the BCS-FFLO phase transition [e.g., along the horizontal dashed line at $\eta=0.25$ in Fig.~\ref{fig:fig4}(b)]. The reason is, in Eq.~\eqref{sus}, the form factor gets Fermi-surface averaged for a dispersive band, which weakens the effects of QGD. And the thermal excitation energy $k_B T$ influences the Fermi-surface average in Eq.~\eqref{sus}, reshaping the boundary line. On the other hand, for conventional superconductors with a highly dispersive band, we have $W \gg U \gg k_B T$ in the weak coupling regime, where the QGD-driven mechanism for the FFLO state may not be effective (more discussions, see SI II~\cite{supp}). Interestingly, previous studies of spin-imbalanced Fermi gases found that the FFLO state becomes unfavorable when the pairing interaction is sufficiently strong~\cite{radzihovsky2010imbalanced}, which stands opposite to the QGD case.

\begin{figure}[t]
        \centering
		\includegraphics[width=1\linewidth]{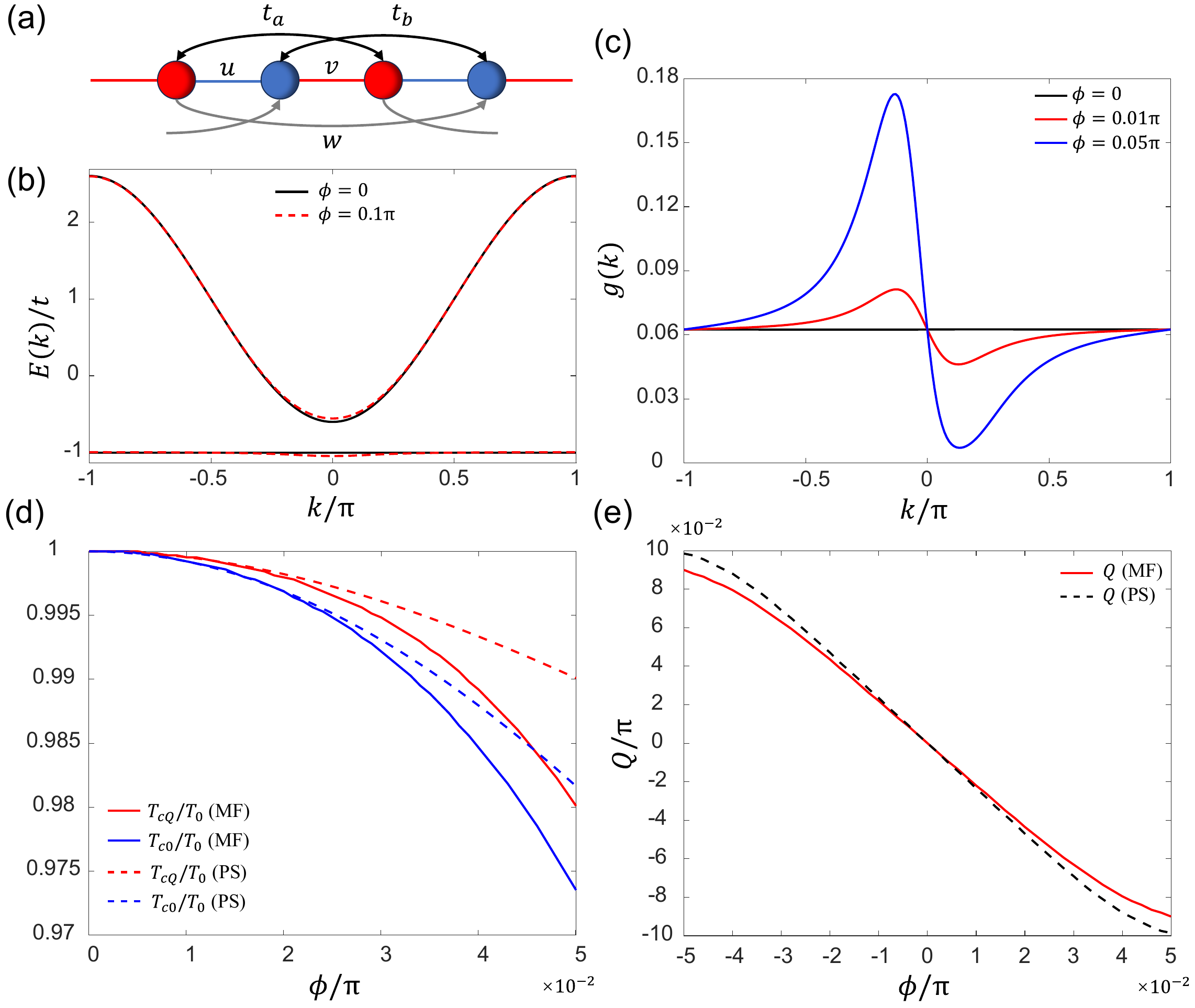}
		\caption{(a) 1D SSH-like model with nearest-neighbor hopping $u$ and $v$, second nearest-neighbor hopping $t_a$ and $t_b$, and third nearest-neighbor hopping $w$. (b) The band structure of the model when $\phi = 0$ (black solid line) and $\phi = 0.1\pi$ (red dashed line). (c) The distributions of quantum metric $g(k)$ at $\phi = 0$, $0.01\pi$, and $0.05\pi$ (labeled in black, red, and blue lines, respectively). (d) The critical temperatures of the most favored FF states and the BCS state are plotted as functions of $\phi$ in red and blue lines, respectively. The results from self-consistent mean-field (MF) calculations are denoted by the solid lines, and the analytical results from Eq.~\ref{critem} (PS) are denoted by the dashed lines. (e) The favored finite-momentum $Q$ with varying $\phi$. Parameters used here: $v = |w| = t_a = t_b = 0.4t, u = -t, U = 0.1t, \nu=0.5.$}
		\label{fig:fig5}
\end{figure}

\section{Flat-band FF state with inversion breaking}
While we did not deliberately distinguish the FF and LO pairings throughout previous sections, we expect the LO-type pairing state to be more stable than the FF type due to the $\pm \bm{Q}$ degeneracy (see Fig.~\ref{fig:fig4}) for the inversion-symmetric model [Eq.~\eqref{eq:hamiltonian}]. Nevertheless, in the following we further demonstrate that if the inversion symmetry is broken in the quantum metric rather than band dispersion, the $\pm \bm{Q}$ degeneracy will be lifted, resulting in the FF-type pairing state driven by QGD.

Here we consider a 1D spinless SSH-like model, as shown in Fig.~\ref{fig:fig5}(a), with the complex nearest-neighbor hopping terms $u$ and $v$, and third nearest-neighbor hopping $w$~\cite{arora2022quantum}. We add the next nearest-neighbor hopping terms $t_a$ and $t_b$ to flatten the lower band [see Fig.~\ref{fig:fig5}(b)], then Hamiltonian can be written as
\begin{equation}\label{SSH}
    H_{1 \mathrm{D}}(k)=
    \left(\begin{array}{cc}
    2 t_a \cos(k) - \mu & J^*(k) \\
    J(k) & 2 t_b \cos(k) - \mu
\end{array}\right),
\end{equation}
where $ J(k) = |u| e^{-i(k+\phi/2)/ 2}+|v| e^{i(k+\phi/2)/ 2}+|w| e^{-i(3(k+\phi/2)/2-\phi)}$, 
and the additional phase shift $\phi/2$ of $k$ is added to make the band dispersion symmetric, i.e., the group velocities satisfy $v_G(\bm{k})=-v_G(-\bm{k})$.

In Fig.~\ref{fig:fig5}(c), we can observe that the quantum metric is asymmetric as $g(k) \neq g(-k)$ when $\phi \neq 0$, which violates the inversion symmetry. At the same time, QGD renders $g_{\uparrow}(k) \neq g_{\downarrow}(-k)$ when we introduce the spin degree of freedom. Therefore, the superconducting BdG Hamiltonian can be expressed as
\begin{equation}
    H_{\mathrm{BdG}}(k,q) = \left(\begin{array}{cc}
         H_{1D}(k+q) & \hat{\Delta}_q\\
        \hat{\Delta}_q^\dagger & 
         -H_{1D}^{T}(-k)
    \end{array}
    \right),
\end{equation}
where $\hat{\Delta}_{q} = \mathrm{diag}[\Delta_{A}(q),\Delta_{B}(q)]$. Following the similar procedures as the previous sections, we can find the critical temperatures as well as the finite-momentum $Q$ of the most favored FF pairing state at the flat band filling, from the self-consistent mean-field (MF) calculations [Fig.~\ref{fig:fig5}(d)], and from the pairing susceptibility (PS) analysis Eq.~\ref{critem} [Fig.~\ref{fig:fig5}(e)]. Interestingly, in Fig.~\ref{fig:fig5}(d), the critical temperature of the most favored FF state will immediately become higher than the BCS state when $\phi$ is turned on. And in Fig.~\ref{fig:fig5}(e), the finite-momentum $Q$ of the most favored FF state is an odd function of $\phi$. Besides, we can see that the mean-field results basically match well with the predictions from Eq.~\ref{critem}, which applies in principle only at the exact flat-band case. Due to the fact that the lower band of this model is not exactly flat when $\phi\neq0$, and becomes more and more dispersive with increasing $\phi$ [see Fig.~\ref{fig:fig5}(b)], the numerical and analytical results slightly deviate from each other when $\phi$ increases. 

The characteristics in Fig.~\ref{fig:fig5}(d) and (e) are quite different from the QGD-driven FFLO phase for the $\zeta$-lattice model in Fig.~\ref{fig:fig3} and Fig.~\ref{fig:fig4}, where the system experiences a first-order BCS-FFLO phase transition driven by QGD. Nevertheless, they mimic the feathers of helical superconductors with spin-orbit interaction~\cite{smidman2017superconductivity,kinnunen2018fulde}. In SI III.B~\cite{supp}, we provide another scheme to realize the flat-band FF pairing state and the emergent superconducting diode effect by turning on an extra inversion-breaking term in Eq.~\eqref{eq:hamiltonian}.

\section{Conclusion and Discussion}
In this work, we have identified the QGD as a generic quantum-geometric origin of the FFLO pairing phase for flat-band electrons, which can be quantified by AQD. While our analysis focuses on the spin-singlet pairing channel, it can be generalized to other pseudo-spin degrees of freedom (such as valley and layer), unconventional pairing channels, and charge as well as spin channels.

Although we adopt an artificial model here to illustrate the effects of the QGD, we emphasize that Eq.~\eqref{sus} and \eqref{critem} provide a general perspective on how the QGD delineates FFLO states. Detailed discussions for additional examples of FFLO states with QGD, e.g., inter-valley pairing between different pseudo-Landau levels, can be found in SI III.C~\cite{supp}. 

\section{Acknowledgements}
We thank Patrick A. Lee, Xiao-Gang Wen, Kun Yang, Xilin Feng, Ying-Ming Xie, Akito Daido, and Xiao Yan Xu for inspiring discussions. K. T. L. acknowledges the support of the Ministry of Science and Technology, China, and Hong Kong Research Grant Council through Grants No. 2020YFA0309600, No. RFS2021-6S03, No. C6025-19G, No. AoE/P-701/20, No. 16310520, No. 16307622, and No. 16309223.

\bibliographystyle{apsrev4-1}
\bibliography{References}

\clearpage
\onecolumngrid
\begin{center}
\textbf{\large Supplemental information for ``Flat-band FFLO State from Quantum Geometric Discrepancy''}\\[.2cm]
 Zi-Ting Sun,$^{1}$  Ruo-Peng Yu,$^{1}$  Shuai A. Chen,$^{2}$ Jin-Xin Hu,$^{3}$  K. T. Law$^{1}$\\[.1cm]
{\itshape ${}^1$Department of Physics, Hong Kong University of Science and Technology, Clear Water Bay, Hong Kong, China
${}^2$Max Planck Institute for the Physics of Complex Systems, N\"{o}thnitzer Stra{\ss}e 38, Dresden 01187, Germany
${}^3$Division of Physics and Applied Physics, School of Physical and Mathematical Sciences, Nanyang Technological University, Singapore 637371}
\end{center}

\maketitle

\setcounter{equation}{0}
\setcounter{section}{0}
\setcounter{figure}{0}
\setcounter{table}{0}
\setcounter{page}{1}
\renewcommand{\theequation}{S\arabic{equation}}

\renewcommand{\thefigure}{S\arabic{figure}}
\renewcommand{\thetable}{S\arabic{table}}
\renewcommand{\tablename}{Supplementary Table}

\renewcommand{\bibnumfmt}[1]{[S#1]}
\renewcommand{\citenumfont}[1]{#1}

\makeatletter
\maketitle


\section{\bf{\uppercase\expandafter{I. Effective action in the band-projection formalism}}}

To study the core physics of the superconducting state in a multi-band system, it is convenient to project the attractive interaction onto the target flat band. In this section, we derive the effective action for the single-band description, with the full numerical results presented in the next section.

\subsection{A. Linearized gap and number equations}
 To begin with, we consider a general local attractive Hubbard interaction:
\begin{equation}
 \hat{H}_{\mathrm{int}}  =-U\sum_{i\alpha}\hat{c}_{i\alpha\uparrow}^{\dagger}\hat{c}_{i\alpha\downarrow}^{\dagger}\hat{c}_{i\alpha\downarrow}\hat{c}_{i\alpha\uparrow}.
\end{equation}
Here $U$ denotes the attractive interaction strength driving the superconducting order. In real space, $i$ and $\alpha$ are the site and orbital index, respectively. The fermion operator $\hat{c}_{\alpha,\bm{k}\sigma}^{\dagger}$ 
in the momentum space can be expressed by Fourier transform via
\begin{equation}
\hat{c}_{\alpha,\bm{k}\sigma}^{\dagger} =\frac{1}{\sqrt{N_c}}\sum_{i}e^{-i\bm{k}\cdot\bm{r}_{i\alpha}}\hat{c}_{i\alpha\sigma}^{\dagger}, \quad
\hat{c}_{i\alpha\sigma}^{\dagger}  =\frac{1}{\sqrt{N_c}}\sum_{\bm{k}}e^{i\bm{k}\cdot\bm{r}_{i\alpha}}\hat{c}_{\alpha,\bm{k}\sigma}^{\dagger}.
\end{equation}
Therefore, we can rewrite the interacting Hamiltonian in the momentum space as
\begin{equation}
 \hat{H}_{\mathrm{int}}=-\frac{U}{N_c} \sum_{\alpha \bm{k}\bm{k}^{\prime} \bm{q}}\hat{c}_{\alpha,\bm{k} +\bm{q} \uparrow}^{\dagger}\hat{c}_{\alpha,-\bm{k} \downarrow}^{\dagger}  \hat{c}_{\alpha,-\bm{k}^{\prime} \downarrow}\hat{c}_{\alpha, \bm{k}^{\prime}+\bm{q}\uparrow}.
\end{equation}
For the convenience of the analytical treatment, we approximate the interaction term~\cite{torma2018quantum,huhtinen2022revisiting,peotta2023quantum} as
\begin{equation}
    \hat{H}_{\mathrm{int}}\approx -\frac{U}{N_{orb}N_c} \sum_{\alpha\beta\bm{k}\bm{k}^{\prime} \bm{q}}\hat{c}_{\alpha,\bm{k} +\bm{q} \uparrow}^{\dagger}\hat{c}_{\alpha,-\bm{k} \downarrow}^{\dagger}  \hat{c}_{\beta,-\bm{k}^{\prime} \downarrow}\hat{c}_{\beta, \bm{k}^{\prime}+\bm{q}\uparrow }.
\end{equation}
Here we have used $g=U/(N_{orb}N_c)$ to characterize the effective interaction. We then project the interaction term onto the target flat band by $\hat{c}_{\alpha, \boldsymbol{k} \sigma} \rightarrow u_{\boldsymbol{k}\sigma}^*(\alpha) \hat{a}_{\boldsymbol{k}\sigma}$, where $\hat{a}_{\boldsymbol{k}\sigma}$ denotes the band-basis electron annihilation operator. Then from $\hat{H}_{\mathrm{int}}$, the effective action $S$ after the band projection can be expressed as
\begin{equation}
  S=\sum_{k, \sigma} \bar{a}_{k \sigma}\left(-i \omega_n+\xi_{\bm{k}\sigma}\right) a_{k \sigma}-g\sum_{q}\bar{\theta}_{q}\theta_{q},
\end{equation}
where $\theta_{q}=\sum_{k}\Lambda^*(\boldsymbol{k},\boldsymbol{q})a_{-k\downarrow}a_{k+q\uparrow}$. Here $\Lambda(\bm{k},\bm{q})=\sum_{\alpha}u_{\boldsymbol{k}+\boldsymbol{q}\uparrow}(\alpha)u_{-\boldsymbol{k}\downarrow}(\alpha)=\langle u_{\bm{k}+\bm{q}\uparrow}|u^*_{-\bm{k}\downarrow}\rangle$ is introduced as the form factor. Here we have used the notations $k=\left(\boldsymbol{k}, \omega_n\right)$ and $q=\left(\boldsymbol{q}, \Omega_m\right)$ with $\omega_n$ $(\Omega_m)$ the fermionic (bosonic) Matsubara frequencies. 

After a standard Hubbard-Stratonovich decoupling, one can find the action in the Nambu basis $\psi_{k,q}=\left(a_{k+q \uparrow}, \bar{a}_{-k \downarrow}\right)^T$ takes the form as
\begin{equation}
   S= \sum_{q} \frac{\left|\Delta_{q}\right|^2}{g}+\sum_{k,q} \bar{\psi}_{k,q} \hat{\mathbb{G}}^{-1}_{k,q} \psi_{k,q},
\end{equation}
where
\begin{equation}
\hat{\mathbb{G}}^{-1}_{k,q} =\left(\begin{array}{cc}
-i \omega_n-i\Omega_m+\xi_{\boldsymbol{k}+\boldsymbol{q} \uparrow} & -\Lambda(\boldsymbol{k},\boldsymbol{q})\Delta_{q} \\
-\Lambda^*(\boldsymbol{k},\boldsymbol{q})\Delta^*_{q} & -i \omega_n-\xi_{-\boldsymbol{k} \downarrow}
\end{array}\right).
\end{equation}
After integrating out the fermionic field, we can get  
\begin{equation}
S[\Delta_q] =\sum_{q} \frac{\left|\Delta_{q}\right|^2}{g}-\operatorname{Tr} \ln \hat{\mathbb{G}}^{-1}_{k,q},
\end{equation}
and expanding it up to the second order in $\Delta_{q}$, the Gaussian action around the trivial saddle point $\Delta_q=0$ reads
\begin{equation}
S_G\left[\Delta_{q}\right]=S[\Delta_{q}=0]+\sum_{q} \Gamma_{q}^{-1}|\Delta_{q}|^2,
\end{equation}
in which the coefficient
\begin{equation}
\Gamma_{q}^{-1}= \left.\frac{\delta^2S[\Delta_{q}]}{\delta \Delta^*_{q}\delta \Delta_{q}} \right|_{\Delta_{q}=\Delta^*_{q}=0}=g^{-1}-\chi^c_{q}
\end{equation}
is the pairing propagator, and the pairing susceptibility reads
\begin{equation}
\chi^c_{q}=\frac{1}{N_c} \sum_{\boldsymbol{k}} \frac{1-n_{\mathrm{F}}\left(\xi_{\boldsymbol{k+q}\uparrow}\right)-n_{\mathrm{F}}\left(\xi_{-\boldsymbol{k}\downarrow}\right)}{\xi_{\boldsymbol{k+q}\uparrow}+\xi_{-\boldsymbol{k}\downarrow}-i\Omega_m}|\Lambda(\boldsymbol{k}, \boldsymbol{q})|^2.
\label{pairsus}
\end{equation}
At the static limit ($\Omega_m=0$), $\chi^c_{\boldsymbol{q}}\equiv \chi^c_{q=(\boldsymbol{q},0)}$, and the linearized gap equation is given by $g\chi^c_{\boldsymbol{q}}=1$.

Near the critical temperature, following the Nosières and Schmitt-Rink (NSR) method \cite{de1993crossover}, we consider the Gaussian fluctuations contribution to the thermodynamic potential as $\Omega=\Omega_0-\beta^{-1} \sum_{q} \ln \Gamma_q$, where $\Omega_0=S[\Delta_{q}=0]/\beta$, and the number equation says $N_{\sigma}=-\partial\Omega/\partial\mu_{\sigma}$. We can rewrite $\Omega$ in terms of a phase shift defined by $\Gamma_{q=(\boldsymbol{q}, i\Omega_m\rightarrow\omega \pm i 0)}=|\Gamma_{q=(\boldsymbol{q}, i\Omega_m\rightarrow\omega )}| \exp [ \pm i \Psi(\boldsymbol{q}, \omega)]$. So the number equation incorporating the effects of Gaussian fluctuations is given by
$N_{\sigma}=N_{0\sigma}+N_{G\sigma}$, where
\begin{gather}
N_{0\sigma}=\frac{1}{2}\sum_{\boldsymbol{k}}\left[1-\tanh \left(\frac{\beta \xi_{\boldsymbol{k}\sigma}}{2}\right)\right],\\
N_{G\sigma}=\sum_{\boldsymbol{q}} \int_{-\infty}^{+\infty} \frac{d \omega}{\pi} n_\mathrm{B}(\omega) \frac{\partial \Psi(\boldsymbol{q}, \omega)   }{\partial \mu_{\sigma}}.
\end{gather}

We can introduce the filling factor $\nu_\sigma=N_\sigma/N_c$. In the ideal flat-band limit, we have $\xi_{\boldsymbol{k}\sigma}=-\mu_\sigma$. When $\nu_{\uparrow}=\nu_{\downarrow}=\nu$ and $\mu_{\uparrow}=\mu_{\downarrow}=\mu$ (no spin-population imbalance), we have $\chi^c_{\boldsymbol{q}}=\tanh \left(\frac{\beta \mu}{2}\right)\frac{1-\bar{d}^2_{\mathcal{A}}(\bm{q})}{2\mu}$. Then we consider the number equation, the free part is $2\nu_0=N_0/N_c=1+\tanh \left(\frac{\beta \mu}{2}\right)$. If we only consider the contribution from $N_0$, i.e., neglecting the fluctuations, we can recover Eq.~(6) in the main text regardless of the filling factor. However, if we include $N_G$, it is not the case for general filling. But worth mention that, Eq.~(6) can approximately stand even with fluctuations when $\beta\mu\ll1$, where $\chi^c_{\boldsymbol{q}}\rightarrow \beta \left({1-\bar{d}^2_{\mathcal{A}}(\bm{q})}\right)/4$. When neglecting $N_G$, this condition corresponds to the neighborhood of half filling.

\subsection{B. Corrections from Gaussian fluctuations}
Then we seriously consider the corrections brought by $N_G$, the dynamical pairing propagator at the ideal flat-band limit is
\begin{equation}
\Gamma_{q}^{-1}=g^{-1}-\tanh \left(\frac{\beta \mu}{2}\right)\frac{1-\bar{d}^2_{\mathcal{A}}(\bm{q})}{2\mu+i\Omega_m}.
\end{equation}
For the convenience of calculation, we notice that
\begin{equation}\label{delta}
\frac{\partial \Psi   }{\partial \mu}=\cos{\Psi}\frac{\partial \sin{\Psi}   }{\partial \mu}-\sin{\Psi}\frac{\partial \cos{\Psi}   }{\partial \mu},
\end{equation}
where $\cos{\Psi}=\mathrm{Re}\Gamma_{q=(\boldsymbol{q}, i\Omega_m\rightarrow\omega \pm i 0)}/|\Gamma_{q=(\boldsymbol{q}, i\Omega_m\rightarrow\omega )}|$ and $\sin{\Psi}=\mathrm{Im}\Gamma_{q=(\boldsymbol{q}, i\Omega_m\rightarrow\omega \pm i 0)}/|\Gamma_{q=(\boldsymbol{q}, i\Omega_m\rightarrow\omega )}|$.

If we replace $i\Omega_m$ by $\omega+ic$, $c$ is an infinitesimal positive number, we can obtain
 \begin{gather}
\cos{\Psi}=\frac{\left(2\mu+\omega\right)^2+c^2 -g \left(2\mu+\omega\right)\left(1-\bar{d}^2_{\mathcal{A}}(\bm{q})\right)\tanh \left(\frac{\beta \mu}{2}\right)}{\sqrt{\left[\left(2\mu+\omega\right)^2+c^2\right]\left[\left(2\mu+\omega-g\left(1-\bar{d}^2_{\mathcal{A}}(\bm{q})\right)\tanh \left(\frac{\beta \mu}{2}\right)\right)^2+c^2\right]}},\\
\sin{\Psi}=\frac{-g c\left(1-\bar{d}^2_{\mathcal{A}}(\bm{q})\right)\tanh \left(\frac{\beta \mu}{2}\right)}{\sqrt{\left[\left(2\mu+\omega\right)^2+c^2\right]\left[\left(2\mu+\omega-g\left(1-\bar{d}^2_{\mathcal{A}}(\bm{q})\right)\tanh \left(\frac{\beta \mu}{2}\right)\right)^2+c^2\right]}}.
\end{gather}
Then from Eq.~\ref{delta} we have
\begin{align}
\frac{\partial \Psi   }{\partial \mu}&=\frac{-g c\left(1-\bar{d}^2_{\mathcal{A}}(\bm{q})\right) \beta}{2 \cosh^2{\left(\frac{\beta \mu}{2}\right)}\left[\left(2\mu+\omega-g\left(1-\bar{d}^2_{\mathcal{A}}(\bm{q})\right)\tanh \left(\frac{\beta \mu}{2}\right)\right)^2+c^2\right]}\\&+\frac{2g c\left(1-\bar{d}^2_{\mathcal{A}}(\bm{q})\right)\tanh \left(\frac{\beta \mu}{2}\right)\left(4\mu+2\omega-g\left(1-\bar{d}^2_{\mathcal{A}}(\bm{q})\right)\tanh \left(\frac{\beta \mu}{2}\right)\right)}{\left[\left(2\mu+\omega\right)^2+c^2\right]\left[\left(2\mu+\omega-g\left(1-\bar{d}^2_{\mathcal{A}}(\bm{q})\right)\tanh \left(\frac{\beta \mu}{2}\right)\right)^2+c^2\right]},    
\end{align}
To evaluate $N_{G}$, we use $\delta(x+a)=\lim_{c\rightarrow0+}\frac{1}{\pi} \frac{c}{(a+x)^2+c^2}$ to 
\begin{equation}
\frac{\partial \Psi   }{\partial \mu}=\left[2\pi-\frac{g \pi\left(1-\bar{d}^2_{\mathcal{A}}(\bm{q})\right) \beta}{2 \cosh^2{\left(\frac{\beta \mu}{2}\right)}}\right]\delta\left(2\mu+\omega-g\left(1-\bar{d}^2_{\mathcal{A}}(\bm{q})\right)\tanh \left(\frac{\beta \mu}{2}\right)\right)
-2\pi\delta\left(2\mu+\omega\right),  
\end{equation}
so that we arrive at
\begin{equation}
N_G=-2N_cn_\mathrm{B}(-2\mu)+\sum_{\bm{q}}\left[2-\frac{g \left(1-\bar{d}^2_{\mathcal{A}}(\bm{q})\right) \beta}{2 \cosh^2{\left(\frac{\beta \mu}{2}\right)}}\right]n_\mathrm{B}\left(-2\mu+g\left(1-\bar{d}^2_{\mathcal{A}}(\bm{q})\right)\tanh \left(\frac{\beta \mu}{2}\right)\right).
\end{equation}

\begin{figure}[t]
        \centering
		\includegraphics[width=0.8\linewidth]{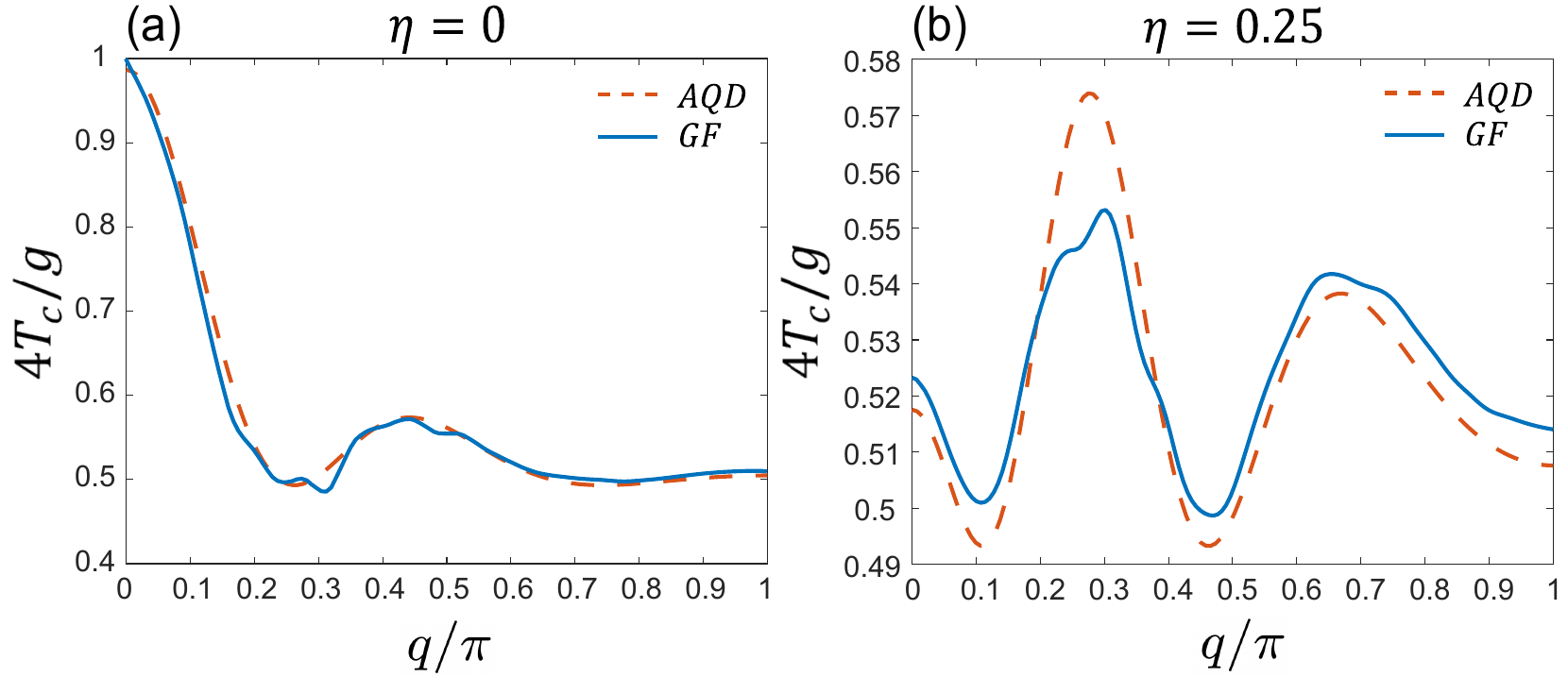}
		\caption{(a) $T_{c\bm{q}}(\eta )$ from Eq.~(6) in the main text (dashed orange line) and from Gaussian fluctuations correction (solid blue line) for $\boldsymbol{q} = q(\hat{x}+\hat{y}),\, q\in \left [ 0,\pi  \right ] $, at $\eta=0$. (b) Same calculations for $\eta=0.25$. In our calculations, we set $\zeta_{\downarrow}=3$ and $\zeta_{\uparrow}=\zeta_{\downarrow}(1+\eta)/(1-\eta)$, doping factor $\delta \nu=0.1$.}
		\label{supfig:supfig1}
\end{figure}

Then we want to show that, even in the presence of Gaussian fluctuations, the FFLO state with the highest critical temperature can still be determined solely by $\bar{d}^2_{\mathcal{A}}(\bm{Q})$. We dope the half filling with $\delta\nu$, then the chemical potential becomes $\mu$. Besides, we use the linearized gap equation concerning the FFLO state with finite momentum $\bm{Q}$, namely, $\beta g=4x/\left[\left(1-\bar{d}^2_{\mathcal{A}}(\bm{Q})\right)\tanh \left(x\right)\right]$ to replace the corresponding $\beta g$ in the number equation, in which we use $x=\beta \mu/2$ for convenience. Then we reduce the number equation to
\begin{equation}
\delta \nu=\frac{1}{2}\tanh{x}-\frac{1}{e^{-4x}-1}+\frac{1}{N_c}\sum_{\bm{q}}\left[1-\frac{2x \left(1-\bar{d}^2_{\mathcal{A}}(\bm{q})\right) }{ \sinh{2x}\left(1-\bar{d}^2_{\mathcal{A}}(\bm{Q})\right)}\right]/ \left[\mathrm{exp}\left(\frac{ 4x\left(\bar{d}^2_{\mathcal{A}}(\bm{Q})-\bar{d}^2_{\mathcal{A}}(\bm{q})\right) }{1-\bar{d}^2_{\mathcal{A}}(\bm{Q})}\right)-1\right].
\end{equation}
We observe that the solution of $x$ is independent of $g$, which implies the critical temperature is always proportional to $g$. When fixing $\delta \nu$, the solution of $x$ is only determined by the value of $\bar{d}^2_{\mathcal{A}}(\bm{Q})$, so does the critical temperature.

However, unlike the free case, $\beta_{c\bm{Q}}$ may not be a monotonic function of $\bar{d}^2_{\mathcal{A}}(\bm{Q})$ now. We can plot $\beta_{c\bm{Q}}$-$\bar{d}^2_{\mathcal{A}}(\bm{Q})$ curve to look for the global minimum point of $\beta_{c\bm{Q}}$ (highest critical temperature), then we can find out the corresponding FFLO instability with $\bar{d}^2_{\mathcal{A}}(\bm{Q})$. We calculate the critical temperatures corrected by Gaussian fluctuations in Fig.~\ref{supfig:supfig1} as a comparison of Fig.~3(b) in the main text. We can see that the emergence of FFLO instability in the presence of Gaussian fluctuations on the whole aligns with the predictions from QGD and AQD.

\section{\bf{\uppercase\expandafter{II. self-consistent multiorbital mean-field theory}}}
In this section, we provide the detailed methodology for the mean-field numerical simulations discussed in the main text. To be specific, we perform the decoupling of the on-site attractive Hubbard interaction within the mean-field approximation:
\begin{equation}
\begin{aligned}     \hat{H}_{\mathrm{int}}&=U\sum_{i\alpha}\hat{c}_{i\alpha\uparrow}^{\dagger}\hat{c}_{i\alpha\downarrow}^{\dagger}\hat{c}_{i\alpha\downarrow}\hat{c}_{i\alpha\uparrow}\\
    & \approx \sum_{\alpha\bm{k}}\left(\Delta_{\alpha}\hat{c}_{\alpha,\bm{k} +\bm{q} \uparrow}^{\dagger}\hat{c}_{\alpha,-\bm{k} \downarrow}^{\dagger} + h.c.\right) + \frac{N_c}{U}\sum_{\alpha}|\Delta_{\alpha}|^2,
\end{aligned}
\end{equation}
where we have used $\Delta_{\alpha}=-U\left\langle\hat{c}_{\alpha,-\bm{k} \downarrow}\hat{c}_{\alpha,\bm{k}+\bm{q} \uparrow}\right\rangle$ and $\Delta_{\alpha} \equiv \Delta_{\alpha}(\bm{q})$ can be determined by the self-consistent gap equation for a given Cooper pair momentum $\bm{q}$. The full mean-field Hamiltonian can be rewritten in the Nambu basis $\psi(\bm{k},\bm{q}) =\left(\hat{c}_{A,\bm{k}+\bm{q}\uparrow}, \hat{c}_{B,\bm{k}+\bm{q}\uparrow},\hat{c}_{A,-\bm{k}\downarrow}^{\dagger},\hat{c}_{B,-\bm{k}\downarrow}^{\dagger}\right)^T$ as
\begin{equation}
\begin{aligned}
    \hat{\mathcal{H}}_{\mathrm{MF}} =\sum_{\bm{k},\bm{q}}\psi^{\dagger}(\bm{k},\bm{q})H_{\mathrm{BdG}}(\bm{k},\bm{q})\psi(\bm{k},\bm{q}) + \sum_{\bm{k},\bm{q}}\operatorname{Tr}[H_{-\bm{k}\downarrow}-\mu(\bm{q})\mathbb{I}_{2\times2}] + \frac{N_c}{U}\sum_{\alpha}|\Delta_{\alpha}|^2,
\end{aligned}
\end{equation}
in which
\begin{equation}
    H_{\mathrm{BdG}}(\bm{k},\bm{q}) = \left(\begin{array}{cc}
         H_{\bm{k}+\bm{q}\uparrow} - \mu(\bm{q})\mathbb{I}_{2\times2} & \operatorname{diag}(\Delta_{\alpha}) \\
         \operatorname{diag}(\Delta_{\alpha}^{\dagger}) & 
         -H_{-\bm{k}\downarrow}^{T} + \mu(\bm{q})\mathbb{I}_{2\times2}
    \end{array}
    \right),
\end{equation}
where $\operatorname{diag}(\Delta_{\alpha}) = \operatorname{diag}(\Delta_A,\Delta_B)$ and the momentum $\bm{q} = q_{x}\hat{x} + q_{y}\hat{y}$.
The grand potential is defined as
\begin{equation}
\begin{aligned}
    \Omega[\mu(\bm{q}),\Delta_{\alpha}(\bm{q}),\bm{q}] & = -\frac{1}{\beta}\operatorname{log}\mathcal{Z} = - \frac{1}{\beta}\operatorname{ln}\operatorname{Tr}[e^{-\beta\hat{\mathcal{H}}_{\mathrm{MF}}}] \\
    & = \sum_{\bm{k}}\operatorname{Tr}(H_{-\bm{k}\downarrow}-\mu(\bm{q})\mathbb{I}_{2\times2}) + \frac{N_c}{U}\sum_{\alpha}|\Delta_{\alpha}|^2 - \frac{1}{\beta}\sum_{\bm{k}}\operatorname{Tr}\left[\operatorname{ln}\left(1 + e^{-\beta H_{\mathrm{BdG}}(\bm{k},\bm{q})}\right)\right] \\
    & = \sum_{\bm{k}}\operatorname{Tr}(H_{-\bm{k}\downarrow}-\mu(\bm{q})\mathbb{I}_{2\times2}) + \frac{N_c}{U}\sum_{\alpha}|\Delta_{\alpha}|^2 - \frac{1}{\beta}\sum_{n\bm{k}}\operatorname{ln}\left(1 + e^{-\beta E_n(\bm{k},\bm{q})}\right),
\end{aligned}
\end{equation}
where $E_n(\bm{k},\bm{q})$ is the n-th eigenvalue of the BdG Hamiltonian $H_{\mathrm{BdG}}(\bm{k},\bm{q})$. By minimizing the grand potential with respect to the pairing potential $\Delta_{\alpha}$ we can get the self-consistent equation:
\begin{equation}
    \begin{aligned}
        & \frac{\partial\Omega} {\partial\Delta_{\alpha}^{*}} = \frac{N_c}{U}\Delta_{\alpha}(\bm{q}) + \sum_{\bm{k}}\langle \hat{c}_{\alpha,-\bm{k} \downarrow}\hat{c}_{\alpha, \bm{k}+\bm{q}\uparrow} \rangle = 0, \\
        & \Delta_{\alpha}(\bm{q}) = - \frac{U}{N_c} \sum_{\bm{k}}\langle \hat{c}_{\alpha,-\bm{k} \downarrow}\hat{c}_{\alpha, \bm{k}+\bm{q}\uparrow} \rangle = - \frac{U}{N_c} \sum_{\bm{k}} \operatorname{Tr}\left[ \mathcal{U}_{\bm{k},\bm{q}}^{\dagger} \mathcal{M}_{\alpha}\mathcal{U}_{\bm{k},\bm{q}}\hat{\rho} \right].
    \end{aligned}
\end{equation}

\begin{figure}[t]
        \centering
		\includegraphics[width=0.95\linewidth]{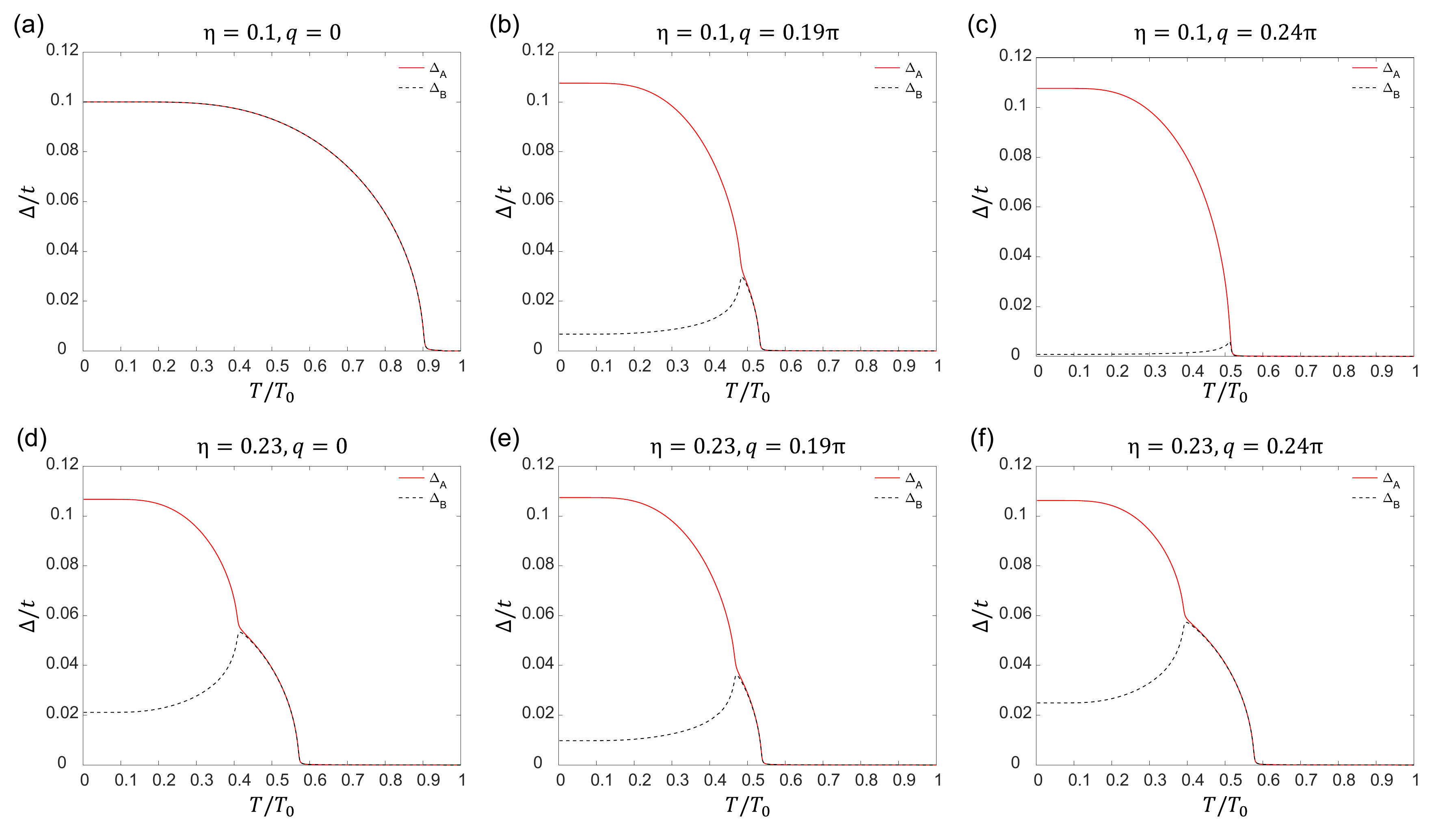}
		\caption{(a)-(c): The $\Delta-T$ relation for the $\eta = 0.1$ (in BCS region) states paired with the finite momenta of $\bm{q} = q(\hat{x}+\hat{y})$, and $q = 0$, $0.19\pi$, $0.24\pi$ respectively. The favoured finite-momentum $Q = 0$. (d)-(f): The $\Delta-T$ relation for the $\eta = 0.23$ (in FFLO region) states paired with the same $q$ in (a)-(c), and the favoured finite-momentum $Q = 0.24\pi$. In (a)-(f), $\Delta_A$ is donated by the red solid curve and $\Delta_B$ is donated by the black dashed curve. In these figures, when $\Delta_A\neq\Delta_B$, we only show the $\Delta_A>\Delta_B$ solution. Of course, there is another solution at $\Delta_B>\Delta_A$ due to the equivalence of $A$, $B$ orbital.}
		\label{supfig:supfig2}
\end{figure}

And the filling factor $\nu$ can be obtained from the thermodynamic relation:
\begin{equation}
    \begin{aligned}
         & N = 2N_c \nu = - \frac{\partial\Omega}{\partial\mu} = \sum_{\bm{k}}\langle \hat{c}_{a,\bm{k}\uparrow}^{\dagger} \hat{c}_{a,\bm{k}\uparrow} +\hat{c}_{a,\bm{k}\downarrow}^{\dagger} \hat{c}_{a,\bm{k}\downarrow} +
        \hat{c}_{b,\bm{k}\uparrow}^{\dagger} \hat{c}_{b,\bm{k}\uparrow} +\hat{c}_{b,\bm{k}\downarrow}^{\dagger}\hat{c}_{b,\bm{k}\downarrow}\rangle, \\
        & 2\nu = 1 + \sum_{\bm{k}} \operatorname{Tr}\left[ \mathcal{U}_{\bm{k},\bm{q}}^{\dagger} \tau_z \otimes 
        \mathbb{I}_{2\times2}
        \mathcal{U}_{\bm{k},\bm{q}} \hat{\rho} \right],
    \end{aligned}
\end{equation}
where $\mathcal{U}_{\bm{k},\bm{q}}$ is the unitary operator that diagonalizes the BdG Hamiltonian as $\mathcal{U}_{\bm{k},\bm{q}}^{\dagger}
H_{\mathrm{BdG}}(\bm{k},\bm{q}) \mathcal{U}_{\bm{k},\bm{q}} = \operatorname{diag}(E_n)$, $\mathcal{M}_A = \left(\begin{array}{cccc}
    0 & 0 & 1 & 0 \\
    0 & 0 & 0 & 0 \\
    1 & 0 & 0 & 0 \\
    0 & 0 & 0 & 0 \\
\end{array} \right)$,
$\mathcal{M}_B = \left(\begin{array}{cccc}
    0 & 0 & 0 & 0 \\
    0 & 0 & 0 & 1 \\
    0 & 0 & 0 & 0 \\
    0 & 1 & 0 & 0 \\
\end{array} \right)$, and the density matrix $\hat{\rho} = \frac{1}{2}\operatorname{diag}\left(\frac{1}{1+e^{\beta E_n(\bm{k},\bm{q})}}\right)$. In turn, if the particle number (filling factor $\nu$) is fixed, the chemical potential $\mu(\bm{q})$ can be determined from this equation. Finally, the free energy density (up to a constant given by the summation $\frac{1}{Nc}\sum_{\bm{k}}\operatorname{Tr}(H_{-\bm{k}\downarrow})$) is:
\begin{equation}
\label{eq:eq_free}
    F(\bm{q}) = \frac{1}{N_c}\left(\Omega[\mu(\bm{q}),\Delta_{\alpha}(\bm{q}),\bm{q}]+\mu(\bm{q})N\right)=\frac{1}{U}\sum_{\alpha}|\Delta_{\alpha}|^2 - \frac{1}{\beta N_c}\sum_{n\bm{k}}\operatorname{ln}\left(1 + e^{-\beta E_n(\bm{k},\bm{q})}\right).
\end{equation}

 We note that in the previous section, the band-projected interaction Hamiltonian is approximated under the uniform pairing condition, where we only consider the trivial irreducible representation as the intra-orbital pairing channel, such that $\Delta_{\alpha \beta}\propto \delta_{{\alpha \beta}}$ is selected as the only relevant pairing channel, while other pairing channels are neglected. To demonstrate that this approximation is valid near the critical temperature in this model, we plot the $\Delta-T$ relation for different values of the QGD $\eta$ and finite momentum $\bm{q}$ in Fig.~\ref{supfig:supfig2}. At low temperatures, $\Delta_A$ and $\Delta_B$ can differ depending on $\eta$ and $\bm{q}$. However, near the critical temperature, we observe that $\Delta_A=\Delta_B$, which satisfies the uniform pairing condition. This observation explains why our critical temperature analysis, based on the uniform pairing condition, aligns well with the numerical results.

\begin{figure}[t]
        \centering
		\includegraphics[width=0.85\linewidth]{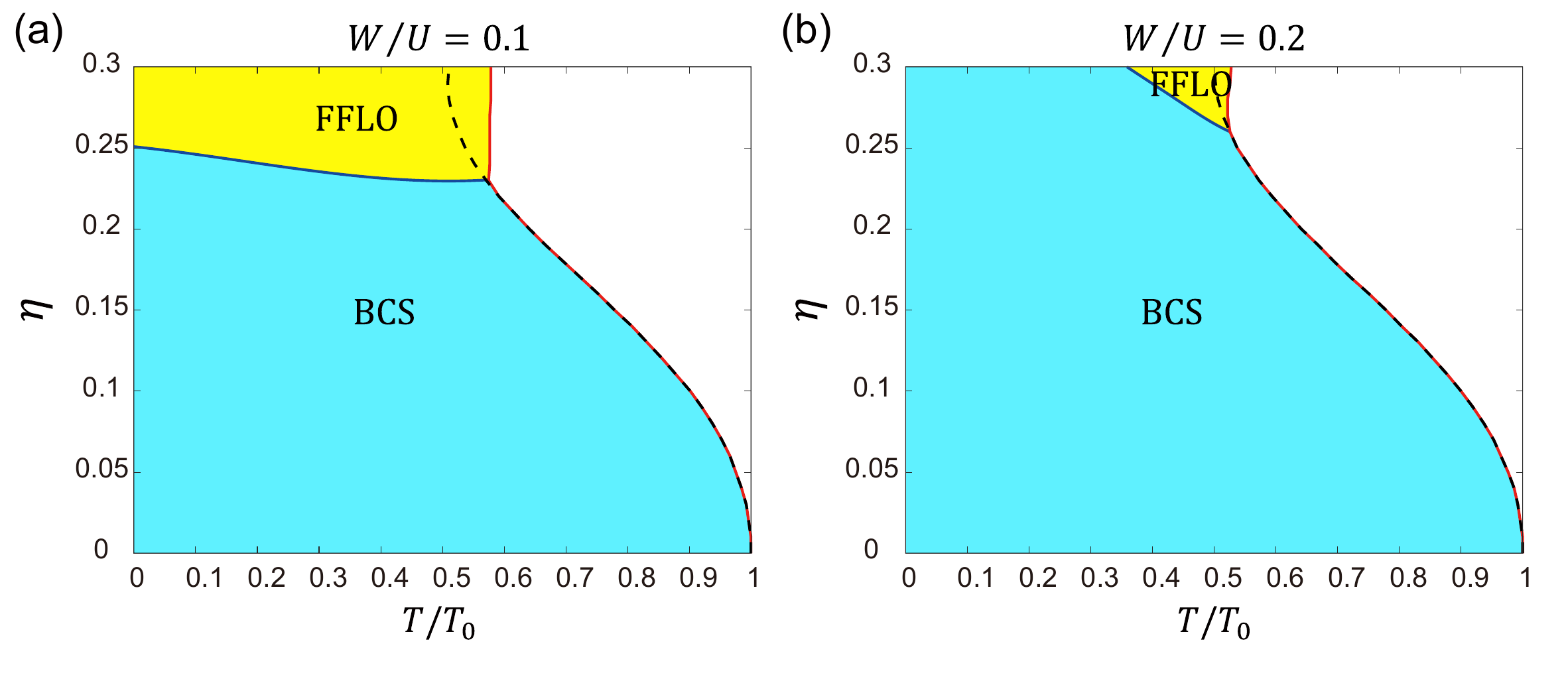}
		\caption{The superconducting phase diagram of the $\zeta$-lattice model. The red solid line labels $T_{c0}(\eta)$, the black dashed line labels $T_{c\boldsymbol{Q}}(\eta)$, where $\boldsymbol{Q} = Q(\hat{x}+\hat{y})$. The BCS regime and the FFLO regime are labeled in light blue and yellow, respectively. Parameters used here: $U = 0.4t$, $\zeta_{\downarrow} = 3$, $W/U=0.1$ in (a) and $W/U=0.2$ in (b).}
		\label{supfig:supfig3}
\end{figure}

We also provide additional numerical results for the case of finite band dispersion. To ensure a narrow bandwidth compared to the band gap and the interaction strength, we maintain the condition $W\ll U\ll 2t$. In Fig.~\ref{supfig:supfig3} we show the phase diagram for $W/U=0.1$ and $W/U=0.2$. Compared to Fig.~4 in the main text, it is evident that the region corresponding to the FFLO phase shrinks as $W/U$ increases. In particular, Fig.~\ref{supfig:supfig3}(b) shows that the FFLO states are confined to a very narrow region. From this trend, it is reasonable to infer that further increasing $W/U$ could lead to the complete disappearance of the FFLO states. This observation supports the claim made in the main text.

\section{\bf{\uppercase\expandafter{III. Additional models for FFLO phase with QGD}}}
\subsection{A. $\zeta$-lattice model on triangular lattice}
\begin{figure}[t]
        \centering
		\includegraphics[width=1.0\linewidth]{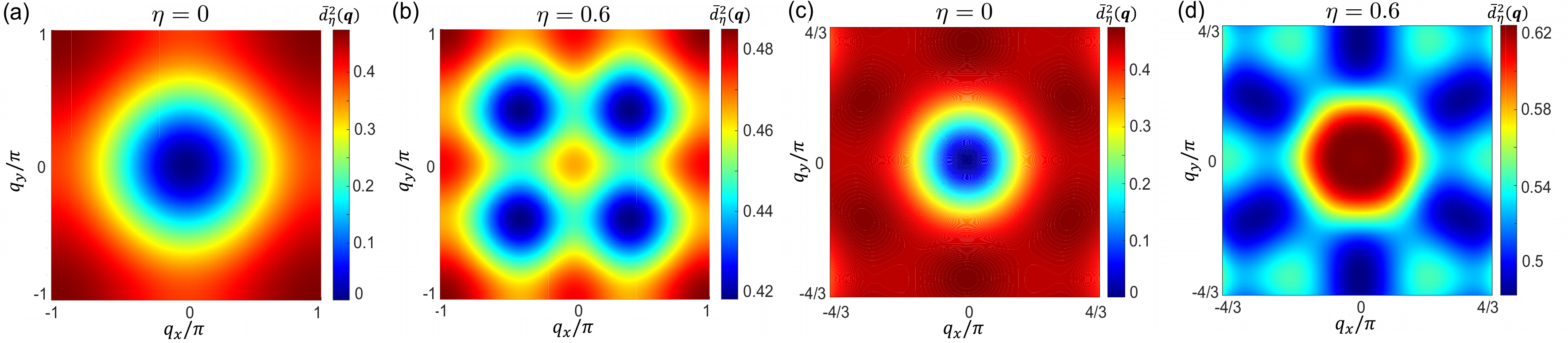}
		\caption{Averaged anomalous quantum distance $\bar{d}^2_{\eta }(\bm{q})$ for the square and triangular $\zeta$-lattice model, in the absence (presence) of QGD. (a) Square lattice for $\eta=0$.    
        (b) Square lattice for $\eta=0.6$. (c) Triangular lattice for $\eta=0$. (d) Triangular lattice for $\eta=0.6$. Other parameters used here: $\zeta_{\uparrow} = 1$, $t = 1$.}
		\label{supfig:supfig4}
\end{figure}

In the main text, we investigate a $\zeta$-lattice model on a square lattice, which preserves $C_4$ symmetry. To further demonstrate the generality of our approach, we extend our analysis in this section to a triangular lattice featuring two orbitals ($A$ and $B)$ per site. Analogously, we can express the Bloch Hamiltonian as follows
\begin{equation}
\label{eq:hamiltonian}
\begin{aligned}
& \mathcal{H}_{\boldsymbol{k} \sigma}=-t \left(\lambda_x \sin \zeta_\sigma p_{\bm{k}}+m_\sigma \lambda_y \cos \zeta_\sigma p_{\bm{k}}\right)-\mu_{\sigma}\lambda_0, \\
& p_{\bm{k}}=\sum_{j=1}^3\cos(\bm{k}\cdot \bm{R}_j).
\end{aligned}
\end{equation}
where $\bm{R}_1=a(1,0), \bm{R}_1=a(-1/2,\sqrt{3}/2),\bm{R}_3=a(-1/2,-\sqrt{3}/2)$. We calculate the averaged anomalous quantum distance $\bar{d}^2_{\eta }(\bm{q})$ for this model in the absence and presence of QGD, as shown in Fig.~\ref{supfig:supfig4}(c) and (d), respectively. For comparison, we also present the corresponding plots for the square lattice model from the main text in Fig.~\ref{supfig:supfig4}(a) and (b). In all panels, the dark blue regions indicate the global minimum of $\bar{d}^2_{\eta }(\bm{q})$. We observe that in both (a) and (c), when $\eta=0$ the global minimum of $\bar{d}^2_{\eta }(\bm{q})$ occurs at $\bm{q}=0$. However, in (b) and (d), the global minima shift away from $\bm{q}$ implying the FFLO instability, as discussed in the main text. Notice that in Fig.~\ref{supfig:supfig4}(b), the FFLO states exhibit four-fold degeneracy, whereas in Fig.~\ref{supfig:supfig4}(d), they display six-fold degeneracy. This difference arises from the distinct crystalline symmetries of the Bloch functions. These results underscore the universality of our theory across various model Hamiltonians.

\subsection{B. Flat-band helical pairing state and superconducting diode effect}
\begin{figure}[t]
        \centering
		\includegraphics[width=1\linewidth]{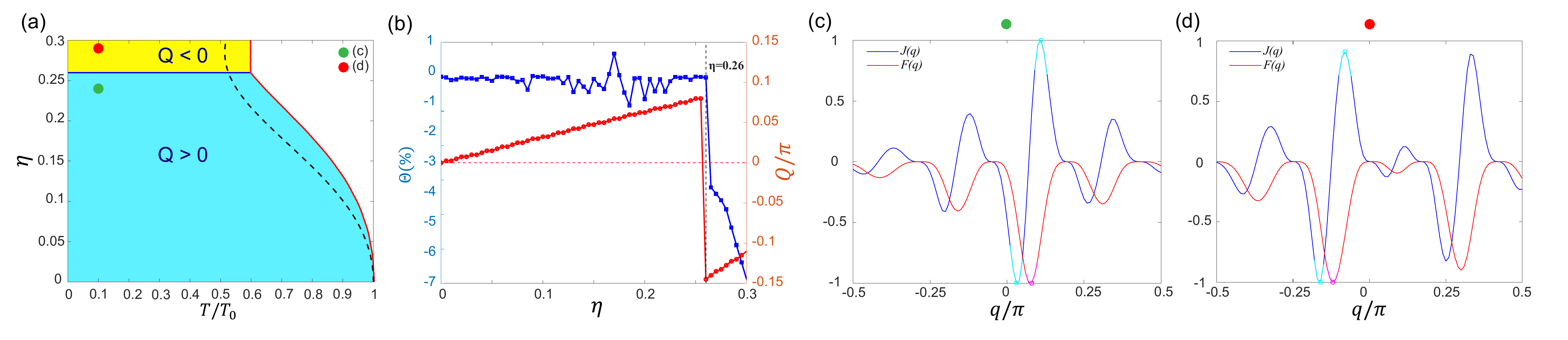}
		\caption{(a) The superconducting phase diagram of $\zeta$-lattice model with an inversion breaking term $\gamma$. The $\boldsymbol{Q} > 0$ (weak helical superconducting state) and $\boldsymbol{Q} < 0$ (strong helical superconducting state) regimes are colored in blue and yellow, respectively. Parameters used here: $U = 0.4t$, $\zeta_{\uparrow} = 3$, $\gamma = 3$.   
        (b) The red curve shows the finite-momentum $\boldsymbol{Q}=Q(1,1)$ versus QGD ($\eta$). Meanwhile, the blue curve is the corresponding diode efficiency $\Theta$. When increasing $\eta$ to across the phase boundary $\eta = 0.26$, the diode efficiency $\Theta$ is enhanced dramatically, and the sign of $\boldsymbol{Q}$ is reversed.
        (c) and (d) The free energy (red curve) and supercurrent (blue curve) of 2 highlighted points in (a) with the temperature $T = 0.1T_0$ but different $\eta$ [$\eta = 0.24$ in (c) and $0.29$ in (d)]. Both the free energy and the supercurrent are normalized to [-1, 1]. The green point is in the $\boldsymbol{Q} > 0$ phase, and the red point is in the $\boldsymbol{Q} < 0$ phase.}
		\label{supfig:supfig5}
\end{figure}

Helical superconductivity can be realized in noncentrosymmetric superconductors with the aid of an in-plane magnetic field~\cite{daido2022intrinsic}. To emulate an analogous helical pairing state in a flat-band system with QGD, we introduce an additional inversion-breaking term manually into the $\zeta$-lattice model. This is done by modifying (recall that the periodic function is $\alpha_{\boldsymbol{k},\sigma}=\zeta_\sigma p_{\bm{k}}$)
\begin{equation}
\alpha_{\boldsymbol{k},\sigma}=\zeta_\sigma p_{\bm{k}} +m_{\sigma}\gamma (\sin k_x +\sin k_y).
\end{equation}
Here the $\gamma$ term breaks the inversion symmetry in the Bloch wave function such that $\alpha_{-\boldsymbol{k},\sigma} \neq \alpha_{\boldsymbol{k},\sigma}$, while still preserving the time-reversal symmetry as $\alpha_{\bm{k},\uparrow} = \alpha_{-\bm{k},\downarrow}$ when $\eta=0$. However, when $\eta\neq0$, this term lifts $\pm\bm{Q}$ degeneracy [see free energy plot in Fig.~\ref{supfig:supfig5}(d)], resulting in FF-type pairing state. Here the Cooper pair momentum is along the $\Gamma-M$ line as $\boldsymbol{q}=q(\hat{x},\hat{y})$.

The mean-field numerical results are summarized in Fig.~\ref{supfig:supfig5}. In this case, the finite momentum pairing state is favored for any nonzero $\eta$. Interestingly, $\boldsymbol{Q}(\eta)$ shows a sharp change near a critical $\eta_c$, manifested as the sign change [see Fig.~\ref{supfig:supfig5}~(b)], which suggests a first-order phase transition. In Fig.~\ref{supfig:supfig5}(a), we color the $\boldsymbol{Q} > 0$ and $\boldsymbol{Q} < 0$ regions in blue and yellow, respectively. This is an analog to the weak and strong helical superconducting states~\cite{daido2022intrinsic}.

Recently, the superconducting diode effect (SDE), in which the magnitudes of the critical supercurrents differ in opposite directions, has been studied in helical superconductors~\cite{daido2022intrinsic,nadeem2023superconducting}. Inspired by this, here we also calculate the diode efficiency $\Theta=(J_{c+}-|J_{c-}|)/(J_{c+}+|J_{c-}|)$ in Fig.~\ref{supfig:supfig5}(b), where $J_{c\pm}$ is the upper and lower critical supercurrent in the $(1, 1)$-direction. It shows that a dramatic enhancement of $\Theta$ appears near $\eta_c$, which is similar to what has been proposed in Ref.~\cite{daido2022intrinsic}. 

The supercurrent can be evaluated by $J(q) = 2[\partial F(q)]/\partial q$, and the minimized free energy can be calculated from Eq.~\eqref{eq:eq_free}.
To give a clearer picture of the phase transition, in Fig.~\ref{supfig:supfig5}(c) and (d), we show the free energy $F(q)$ (red curves) as well as the supercurrent $J(q)$ (blue curves) for the two points (green and red dots) in Fig.~\ref{supfig:supfig5}(a). In the blue region [Fig.~\ref{supfig:supfig5}~(a)], the global minimum of the free energy $F(q)$ is at $Q>0$ [Fig.~\ref{supfig:supfig5}~(c)], but $|\bm{Q}|$ is small, associated with the small diode efficiency. When $\eta >\eta_c$ [Fig.~\ref{supfig:supfig5}~(d)], the pairing state with $Q < 0$ becomes the global minimum. In this case, an enhanced diode efficiency can be observed in Fig.~\ref{supfig:supfig5}(b).

\subsection{C. Inter-flavor FFLO phase within pseudo-Landau levels}
\begin{figure}
    \centering
    \includegraphics[width=1\linewidth]{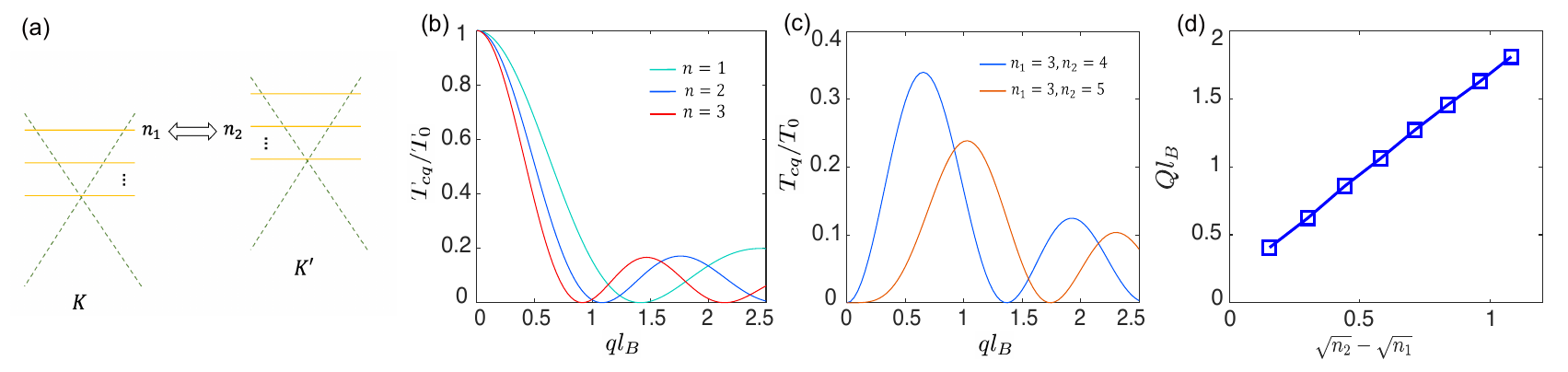}
    \caption{(a) The inter-flavor (valley) pairing involves the electrons on Landau level $n_1$ from $K$ valley and $n_2$ from $K'$ valley. (b) The $T_{cq}$ for the BCS-type pairing ($Q=0$) at $n_1=n_2=n$. (c) Calculated $T_{cq}/T_0$ for the FFLO pairng at $n_1\neq n_2$. (d) The stablized finite momentum $Q$ scales as $\sqrt{n_2}-\sqrt{n_1}$ at $n_1=10$.}
    \label{supfig:supfig6}
\end{figure}

In this section, we explore the formation of the FFLO phase in pseudo-Landau level systems. As schematically illustrated in Fig.~\ref{supfig:supfig6}, we consider a strained spinless Dirac material with a series of pseudo-Landau levels under a uniform pseudomagnetic field.  The $K$ and $K'$ valleys are time-reversal partners, which naturally leads to the inter-flavor Cooper pairings in the presence of attractive interactions. Under this scenario, we consider Cooper pairs form within Landau level $n_1$ from $K$ valley and $n_2$ from $K'$ valley. Here, an imbalance in the populations of the two flavors (such as valley polarization), i.e., $\mu_{K}\neq \mu_{K'}$ may lead to $n_1\neq n_2$, where QGD comes into being as the quantum metrics of the different pseudo-Landau levels are distinct, and the FFLO phase can manifest. This scenario is also proposed in Skyrmion crystals~\cite{dong2024chirality}.

To evaluate the critical temperature $T_{cq}$, we first use the form factor of Landau levels on the torus geometry to obtain the anomalous quantum distance~\cite{liu2024theory}:
\begin{equation}
\langle u_{n_2}^K(\bm{k}+\bm{q})|u_{n_1}^{*K'}(-\bm{k})\rangle =\langle u_{n_2}^K(\bm{k}+\bm{q})|u_{n_1}^{K}(\bm{k})\rangle=e^{\frac{i}{2}(\bm{k}+\bm{q})\times \bm{k}}\sqrt{\frac{n_1 !}{n_2 !}}L_{n_1}^{n_2-n_1}(qq^*)(iq)^{n_2-n_1}e^{-\frac{1}{2}qq^*}.
\end{equation}
Here $q=(q_x+q_y)/\sqrt{2}$ and $L_{n}^m (x)$ is the generalized Laguerre polynomial. And we have used $|u_{n_1}^{*K'}(-\bm{k})\rangle =|u_{n_1}^{K}(\bm{k})\rangle$. Then we use the formula in main text $T_{c\boldsymbol{q}}/T_0=1-\bar{d}^2_{\mathcal{A}}(\boldsymbol{q})$ to calculate $T_{c\boldsymbol{q}}$ as summarized in Fig.~\ref{supfig:supfig6}.

Firstly, we notice that the zero-momentum pairing is favored when $n_1=n_2=n$ as shown in Fig.~\ref{supfig:supfig6}(b). Notice that for Landau levels the quantum metric satisfies $\int tr[g_n(\bm{k})]=2n+1$. Thus $T_{cq}/T_0=1-\frac{2n+1}{2}q^2$ for small $q$. In Fig.~\ref{supfig:supfig6}(c), when $n_1 \neq n_2$, it can be seen that a global maximum occurs at $q=Q$ corresponding to the FFLO phase. 

We can make an rough estimation of the finite momentum $Q$ by evaluating $\partial  [L_{n_1}^{n_2-n_1}(q^2/2)(q/\sqrt{2})^{n_2-n_1}]/\partial_q=0$ at $q=Q$, leading to
\begin{equation}
-Q^2 L_{n_1-1}^{n_2-n_1+1}(Q^2/2)+(n_2-n_1)L_{n_1}^{n_2-n_1}(Q^2/2)=0.
\end{equation}
Expanding it to the lowest order of $Q$ and at large $n_1$, $n_2$, we can make an approximate result as
\begin{equation}
Q\approx \sqrt{\frac{2(n_2-n_1)\binom{n_2}{n_1}}{(n_2-n_1+2)\binom{n_2}{n_1-1}}}=\sqrt{\frac{2(n_2-n_1)(n_2-n_1+1)}{n_1(n_2-n_1+2)}}.
\end{equation}
For $n_2$ close to $n_1$, $Q\sim \sqrt{n_2}-\sqrt{n_1}$ as demonstrate in Fig.~\ref{supfig:supfig6}(d). We can also observe a direct relationship between 
$Q$ and the QGD, as the quantum metrics of the different Landau levels are distinct.

\end{document}